\documentclass[12pt]{article}
\usepackage{epsfig}
\usepackage[hang,bf,small]{caption}
\DeclareGraphicsExtensions{.eps.gz,.eps,.ps,.ps.gz}
\parskip=\itemsep     
\begin{document}
\title{
{\Large \bf   Saturation at low $\mathbf{x}$ }}
\author{
{\large  ~ E.~Levin,\thanks{e-mail: leving@post.tau.ac.il}
} \\[4.5ex]
{\it  HEP Department}\\
{\it  School of Physics and Astronomy}\\
{\it Raymond and Beverly Sackler Faculty of Exact Science}\\
{\it Tel Aviv University, Tel Aviv, 69978, ISRAEL}\\[4.5ex]
}

\maketitle
\thispagestyle{empty}
                      
\begin{abstract} 

This talk is an attempt to review  all our knowledge on  saturation at low
$x$ both theoretical and experimental, to stimulate a search for saturation
effects at THERA.  The main goals of this presentation are

1. \,\, To discuss an intuitive picture of the deep inelastic scattering
that leads   to the saturation of the parton densities;

2. \,\, To show that the saturation hypothesis has solid theoretical
proof;

3. \,\,  To report on  the  theoretical progress that has been made over
  the past two years in high parton density QCD, and on the property of
  the saturation phase that emerges from the  theory that has been developed;

4. \,\,  To collect all    that  we know  theoretically and
   experimentally  about the saturation   scale $Q_s(x)$ .

 \end{abstract}

\begin{flushright}
\vspace{-16.5cm}
TAUP - 2681-2001 \\
\today
\end{flushright}   
\thispagestyle{empty}  
\newpage

\setcounter{page}{1}

\begin{flushright}
{\large \bf \it  Fighting  prejudices}
\end{flushright}
\section{Introduction}
\begin{figure}[h]
\begin{center}
  \epsfxsize=12cm
\leavevmode
\hbox{ \epsffile{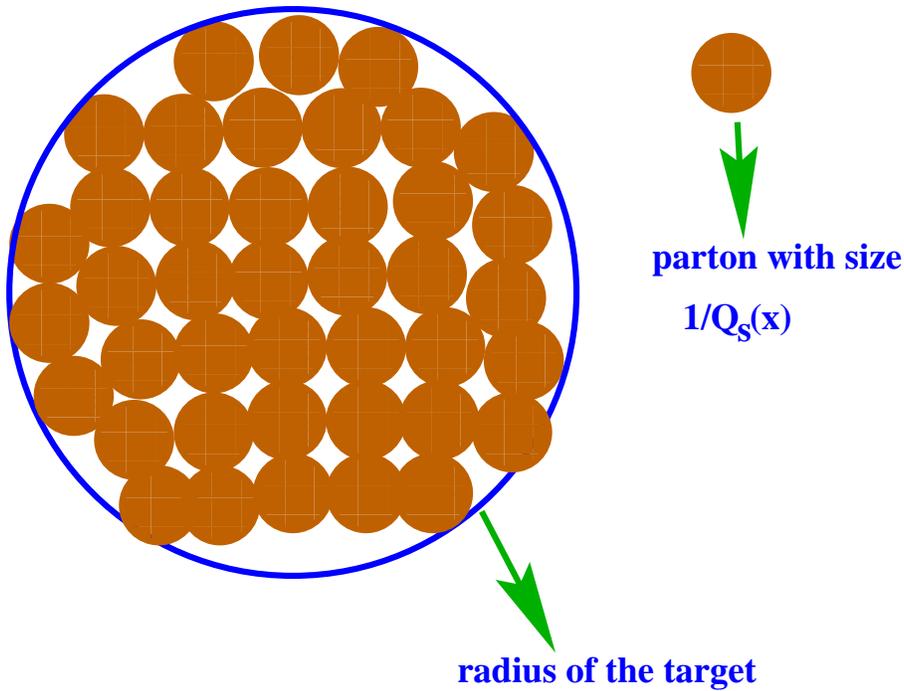}}
\end{center}
\caption{Picture of a hadron in the saturation region.}
\label{fig:lev:sat}
\end{figure}

At low values of $x$,  QCD evolution, both  DGLAP\cite{DGLAP} and 
BFKL\cite{BFKL} ,  predict a striking increase of the parton densities
which violate  unitarity constraints
\cite{GLR}. Therefore,  interactions between partons in the parton cascade,
omitted in QCD evolution equations, should  become  essential to slow down
the growth of the parton densities. We expect that these interactions will
create an  equilibrium -like system of partons with a definite value for  the
average transverse momentum, which we call a saturation scale ($Q_s(x)$). In
other words,  we expect a picture of a  hadron as shown in Fig.~\ref{fig:lev:sat}\cite{GLR,MUQI,MV,MU99}.

This talk is an attempt to review  all our knowledge on  saturation at low
$x$ both theoretical and experimental, to stimulate a search for saturation
effects at THERA.  In spite of the fact that the high density QCD phase was
only briefly discussed  in the  THERA Contribution to the TESLA TDR, everybody
knows that {\bf if  saturation effects are  seen  at THERA it will  be the
  machine, while if not,  it will remain  one of many}. 
The main goals of this presentation are
\begin{itemize}
\item \quad To discuss an intuitive picture of the deep inelastic scattering 
that leads
  to the saturation of the parton densities;
\item \quad To show that the saturation hypothesis has solid theoretical 
  proof;
\item \quad To report on  the  theoretical progress that has been made over
  the past two years in high parton density QCD, and on the property of
  the saturation phase that emerges from the theory that has been developed;
 \item \quad  To collect all   that  we know  theoretically and
   experimentally about the saturation   scale $Q_s(x)$.
\end{itemize}

Thera are  two different ways to reach a high parton density phase:
the first, is DIS at low $x$, the second is  deep inelastic scattering on  a
nuclear target in which we have a rather large parton density from the
beginning,  due to a  large number of nucleons in a nucleus. The best
avenue to investigate the high density phase is to use both paths  and to measure
DIS on  nuclei at low $x$. This is one of the THERA options and we will
also discuss  saturation phenomena for such a reaction.

\section{Qualitative Picture of Interaction in DIS at low $\mathbf{x}$}

\subsection{Bjorken frame:}

\begin{figure}[h]
\begin{center}
  \epsfxsize=16cm
\leavevmode
\hbox{ \epsffile{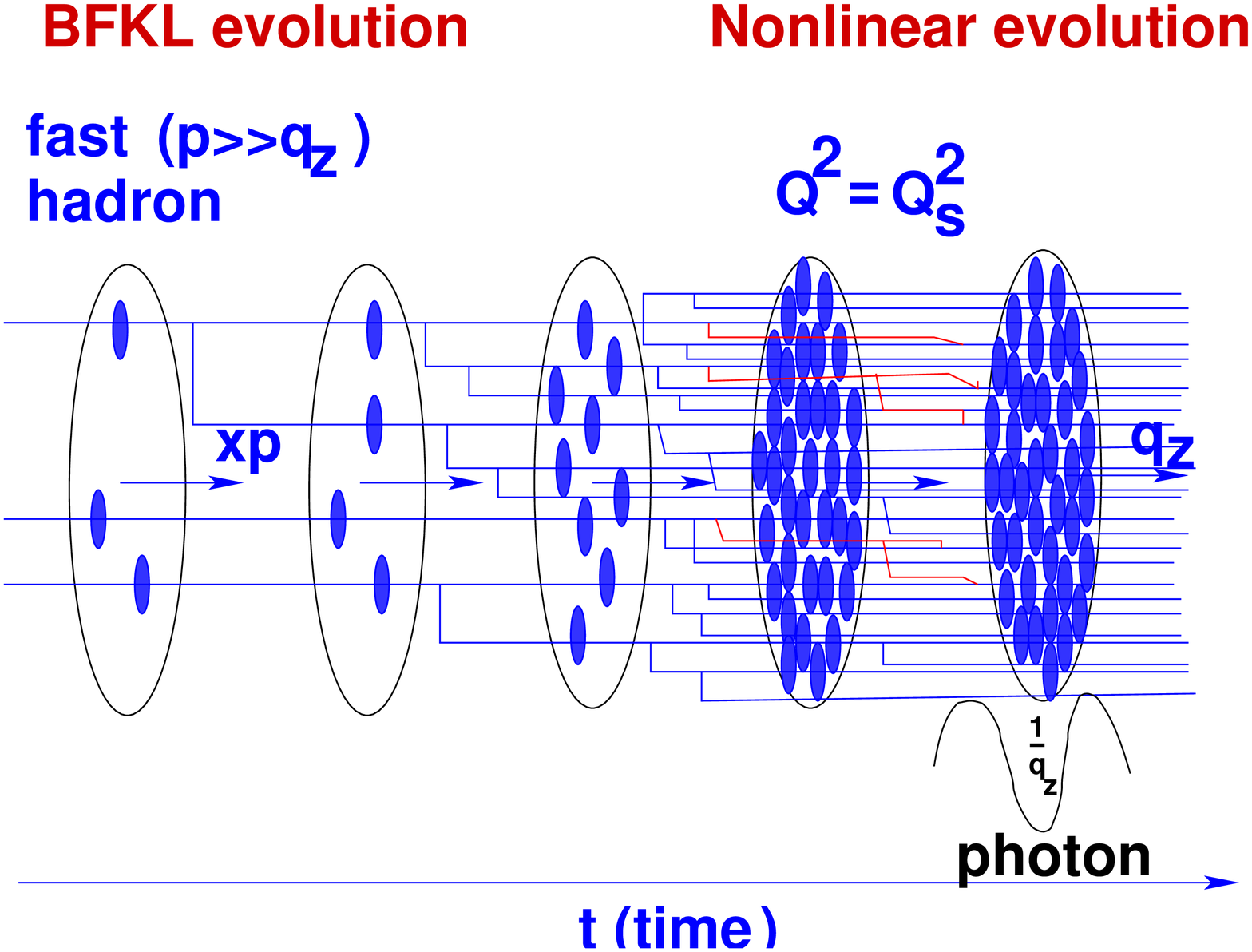}}
\end{center}
\caption{Parton cascade in the Bjorken frame.}
\label{fig:lev:bjfr}
\end{figure}

It is well known that deep inelastic scattering is most clearly visualized in
a 
space-time picture in the Bjorken frame where the virtual photon has zero
energy \cite{BJ}. Therefore, in the Bjorken frame the electro-magnetic field
is a standing wave with wavelength of the order of $1/q_z$ for a  photon
with four momentum $q_{\mu} = (0, q_z,0,0,)$ (see Fig. ~\ref{fig:lev:bjfr}). In this frame the fast hadron
decays into  a system of partons. Each parton has a longitudinal  momentum
$p_{i,z} = x_i P$, where $x_i$ is a fraction of the energy of incoming hadron
carried by the parton, and a transverse momentum $p_{i,t}$. Due to the
uncertainty principle each parton is localized in $\Delta z_i \approx 1/(x_i
P)$ and, therefore, only partons with $x_i P \approx q_z$ can interact with
the photon, since for all other partons the overlap  integral is very small. 
In other words, the parton which interacts with the photon has $x_i \approx
q_z/P $. Using the energy and momentum conservation for the parton - photon
interaction one can easily obtain\footnote{Energy conservation gives that the
  energy of the parton $i$ and the recoiled energy are equal\,\,\,\,($E_i = E'_i$) , the conservation
  of the longitudinal moment  leads to  $p_{i,L} = q_z - p'_{i,L}$. $p'_{i,L}
  = p_{i,L}$ and $p_{i,L} = q_z/2$.}
that $x_i P = q_z/2$ which gives $x_i = q_z/2P = q^2_z/2 P q_z = Q^2/2(P\cdot
q) = Q^2/s = x_{Bj}$ at low $x$.

The lifetime of the $i$-th parton is of the order of $\tau =  \tau_{r.f.}
\gamma  =\frac{
  E_i}{p^2_{i,t}} = \frac{x_i P}{p^2_{i,t}}$ where the lifetime in the rest
frame of the $i$-parton $
\tau_{r.f.} = \frac{1}{p_{i,t}}$ and $\gamma = \frac{E_i}{p_{i,t}}$. The
parton that  interacts with the photon lives for  a short time $ \approx 1/q_z$
and because of this the interaction cannot change the parton distribution, and
destroys only the coherence of the partons in the incoming hadron.

Over a  long period of time every parton can decay into a large number of
partons. Theoretically we can  only control the emission of the
partons with large values of the transverse momenta, as  the QCD coupling
is small for them, and we can safely use the developed methods of perturbative
QCD.  Fig. ~\ref{fig:lev:bjfr} shows  the evolution of the partons
with definite transverse momenta ($ p_t = Q \gg 1/R$) (with definite size $r_t =
1/Q \ll R$) in time ( so called the BFKL evolution). $R$ is the size of the
hadron. The first stage of the evolution for partons with $x_i \approx 1$ is
not under   theoretical control and only non-perturbative QCD will be able to
give us  information on probability   ($P^i_h(x_i,p_{i,t} = Q \ll 1/R$ ) to find
several partons
 with  $ p_{i,t} = Q $ and $x_i \approx 1$ in the hadron.  The BFKL evolution
 takes into account the emission of  partons with $ p_t = Q \gg
 1/R$. Fig. ~\ref{fig:lev:bjfr} shows that it is natural to expect that
 this emission leads to a  considerable increase of the number of 
 partons.  The number of partons that can interact with the target ( virtual photon)
can be written in the form  of the  convolution $F^{BFKL}(\frac{x}{x_i},Q^2)
\,\bigotimes\, P^I_h(x_i,Q)$. We can obtain the result of the DIS experiment 
by taking the convolution (overlapping integral ) with the photon wave
function. In other words, the deep inelastic structure function is equal to
\begin{equation}\label{eq:lev:f2}
  F_2 (x, Q^2) =
  P_{\gamma^*}(\frac{x_{Bj}}{x},Q)\,\bigotimes\,F^{BFKL}(\frac{x}{x_i},Q^2)
\,\bigotimes\, P^I_h(x_i,Q)\,\,,
\end{equation}
where $  P_{\gamma^*}(\frac{x_{Bj}}{x},Q) = |\Psi_{\gamma^*}(
\frac{x_{Bj}}{x},Q)|^2$  and   $\Psi_{\gamma^*}$ is the wave function of the
virtual photon.

Recalling that $\sigma( \gamma^*, h) = \frac{4 \pi^2}{Q^2}
\,\,F_2(x_{Bj},Q^2)$ one can see that the unitarity constraint $\sigma(
\gamma^*, h) \,\leq\,\pi R^2$ leads to a conclusion that the increase of the
parton densities due to the BFKL ( or DGLAP) emission should be tamed \cite{GLR}. The
simple idea how such taming can occur is clear from
Fig. ~\ref{fig:lev:bjfr}. Indeed, if the parton cascade has been measured at
early time ( at rather high $x$ )  the density of the partons in the
transverse plane (see Fig. ~\ref{fig:lev:bjfr}) is not large, and
 we have to take into account emission,   since emission is proportional to the
density( $\rho$) of partons ( {\bf emission $\mathbf{\propto}$  $
  \mathbf{\rho}$}). However, the density of partons increase due to emission
and at some  value of $x$ the system becomes so dense that  partons cover the
hadron disc. In such a situation the interactions of the partons start to be
essential. These interactions are proportional  to the square of the parton
density since two partons have to meet in one point for this interaction (
{\bf annihilation }$\mathbf{\propto}$ \,$\mathbf{(\alpha_S/Q^2)}\,\cdot\,$
  $\mathbf{\rho^2}$,
where $\alpha_S/Q^2$ is the typical cross section of
two parton annihilation in the parton cascade) 
 and they
cause  the number of particles to diminish. Therefore, we expect there to be  an
equilibrium between emission and annihilation in the dense system of partons
which  can be described by simple equation:

\begin{equation}\label{eq:lev:glrsim}
\frac{d \rho}{d \ln(1/x)\,}\,\,\,=\,\,\frac{N_c\,\alpha_S}{\pi}\,\,\left(
  K^{BFKL}\,\bigotimes\, \rho\,\,\,-\,\,\frac{\gamma\,\alpha_S}{Q^2}\times\rho^2
  \,\right)\,\,.
\end{equation}
The first term of this equation gives the BFKL evolution at low $x$ while
the second one provides a taming of the density increase. Of course, all
coefficients in 
 the equation cannot be calculated in framework of such
oversimplified approach, including numerical coefficient $\gamma$.

\begin{figure}[h]
\begin{center}
  \epsfxsize=12cm
\epsfysize=12cm
\leavevmode
\hbox{ \epsffile{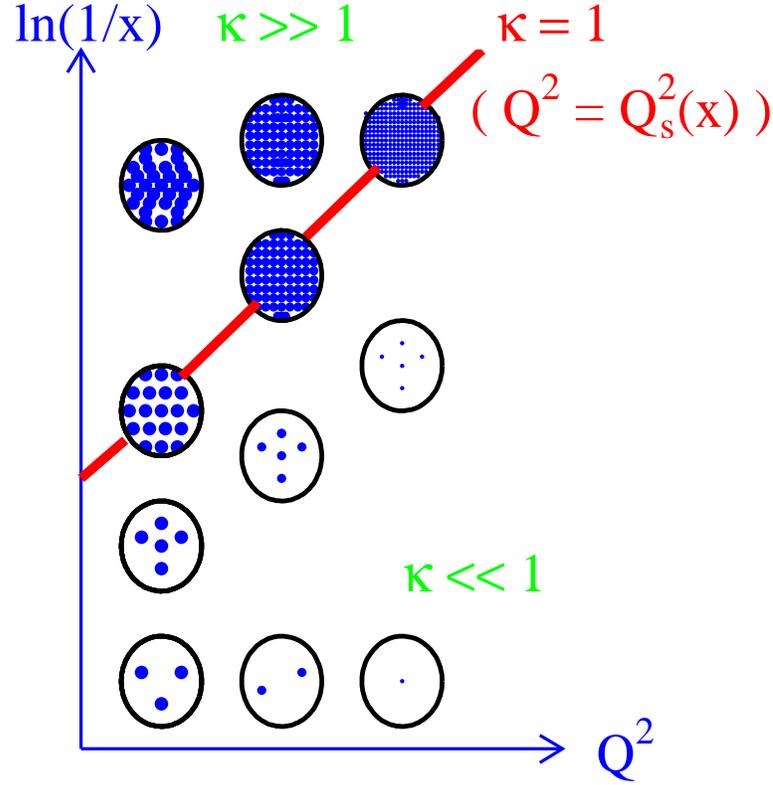}}
\end{center}
\caption{Parton distribution in transverse plane and saturation  in the Bjorken frame.}
\label{fig:lev:satpic}
\end{figure}

The principle prediction of this equation is the saturation of the parton
density, namely, the fact that the parton density stops   increasing.  It
should be stressed
that at any value of
$Q^2$, even a very large value, there exists a small value of $x$ at which we
face  saturation (  see Fig.~\ref{fig:lev:satpic} ). $\kappa \,\propto\,\rho$ in
Fig.~\ref{fig:lev:satpic} is a packing factor for the partons and we will discuss its value
 later.

\subsection{Laboratory frame:}
\begin{figure}[h]
\begin{center}
  \epsfxsize=16cm
\leavevmode
\hbox{ \epsffile{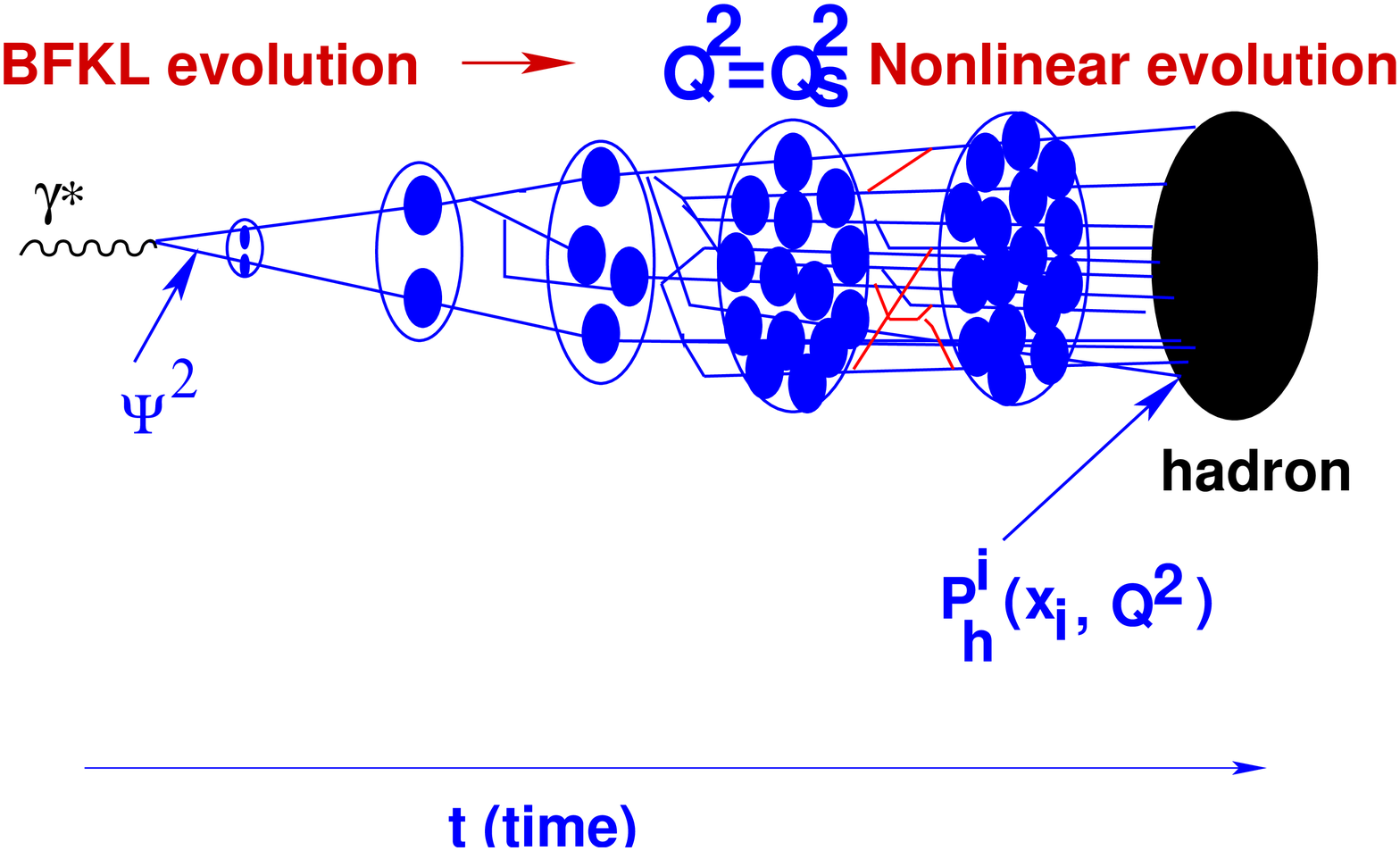}}
\end{center}
\caption{Parton cascade in the laboratory frame.}
\label{fig:lev:labfr}
\end{figure}

The Bjorken frame is the frame which is best suited for the discussion of DIS  in the
parton (QCD) approach, since both pictures  Fig. ~\ref{fig:lev:bjfr} and   Fig.~\ref{fig:lev:satpic}
give the parton distributions in a hadron, or, in other words, term $,F^{BFKL}(\frac{x}{x_i},Q^2)
\,\bigotimes\, P^I_h(x_i,Q)$  in Eq.~\ref{eq:lev:f2}. However, it turns out
that some properties of the high density parton system are  clearer in
the laboratory frame where the hadron is at  rest. As we will see below,
the fact that our partons are colour dipoles is easy to
demonstrate in this frame.
 The time-space picture
of  DIS in this frame is shown in Fig.~\ref{fig:lev:labfr}.

In the lab. frame the fast virtual photon decays into a  quark-antiquark pair ( two
partons in Fig.~\ref{fig:lev:labfr}). Quark ( antiquark)  has transverse momentum larger
than $Q$ and  exists for a   sufficiently long time ($ \tau
\,\approx\,1/mx$)\footnote{The estimates for $\tau$ we can easily obtain
  using the uncertainty principle:$ \Delta E \,\tau \,\approx\,1$ where
  $\Delta E$ is the difference in energy between initial and final
  states. For the virtual photon decay we have $  \Delta E = q_0 - p_{1,0} +
  p_{2 ,0} \approx q_0 - q_z = mx$, where $p_{i,0}$ is the energy of produced
  quark (antiquark).}. If the time $\tau$  is long enough,  quarks
(antiquarks) radiate gluons ( as
shown in Fig.~\ref{fig:lev:labfr})  which   create a dense
 parton system   in the same way as in the Bjorken frame. 

At first sight pictures  Fig. ~\ref{fig:lev:bjfr} and
Fig.~\ref{fig:lev:labfr} look quite different. Of course, the final result of
the measurement (the total photon-hadron cross section) given by Eq.~\ref{eq:lev:f2} , remains the same in
both frames, but only part of Eq.~\ref{eq:lev:f2} is shown  in
Fig.~\ref{fig:lev:labfr}, namely, $
P_{\gamma^*}(\frac{x_{Bj}}{x},Q)\,\bigotimes\,F^{BFKL}(\frac{x}{x_i},Q^2)$.

Therefore, the main difference between these two figures,
Fig. ~\ref{fig:lev:bjfr} and Fig.~\ref{fig:lev:labfr} is the following:
Fig. ~\ref{fig:lev:bjfr} shows all partons which can interact with the photon
target,  while the system of partons that has interacted with the virtual
photon is depicted in Fig.~\ref{fig:lev:labfr}. One can see that
the Bjorken frame is much better for describing  the parton  densities,  we
will show  in the next section  the laboratory frame is very useful in
answering  the question: What are these partons in QCD.

\subsection{Colour Dipoles = Partons:}

\begin{figure}[h]
\begin{center}
  \epsfxsize=15cm
\leavevmode
\hbox{ \epsffile{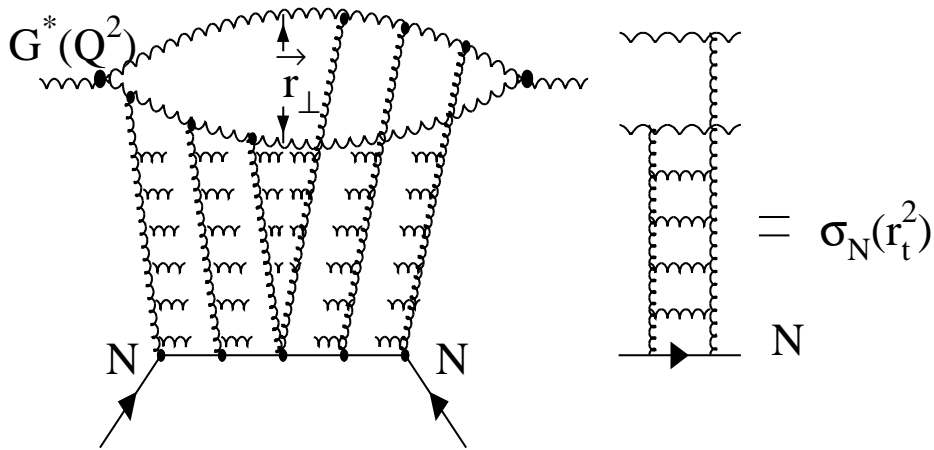}}
\end{center}
\caption{Interaction of a colour dipole with the target (Glauber-Mueller formula) in the laboratory frame.}
\label{fig:lev:mgfor}
\end{figure}

The advantage of the lab. frame becomes clear if we want to understand how
the 
produced quark -antiquark pair( which is a colour dipole)  interacts with the
target. The main observation is that the size ($r_{\perp}$ in Fig. ~\ref{fig:lev:mgfor})
of the  colour dipole or , in other words, the transverse distance between the
quark and antiquark, is a good degree of freedom, which is preserved by the high
energy QCD interaction \cite{DOF1,DOF2,DOF3}. Indeed, while the colour dipole
is traversing  the target, the distance $r_{\perp}$ between the quark and
antiquark can vary by an amount $\Delta  r_{\perp} \,\propto\,R
\,\frac{k_{\perp}}{E}$, where $E$ denotes the energy of the dipole in the
lab. frame and $R$ is the size of the target (see  Fig. ~\ref{fig:lev:mgfor}
). Due to the uncertainty principle the quark transverse momentum is
$k_{\perp}\,\propto\, \frac{1}{ r_{\perp}}$. Therefore, 

\begin{equation}\label{eq:lev:deltar}
\Delta r_{\perp} \,\,\propto\,\, R \,\frac{k_{\perp}}{E}\,\,\approx\,\,R
\,\frac{1}{r_{\perp}\,E}\,
\ll\,\,r_{\perp}\,\,,
\end{equation}
if $$
r^2_{\perp}\,2 m E  \,\gg\,2\,m\,R. $$
Since $r^2_{\perp} \,\approx \,1/Q^2$,  and recalling the definition of the
Bjorken $x$ one can see that 
\begin{equation}\label{eq:lev:drr}
\frac{\Delta r_{\perp}}{r_{\perp}}\,\,\ll\,\,1 \,\,\,\,\,
\mbox{at}\,\,\,\,\, x \,\,\ll\,\, \frac{1}{2 m R}.
\end{equation}

\begin{figure}[h]
\begin{center}
  \epsfxsize=15cm
\leavevmode
\hbox{ \epsffile{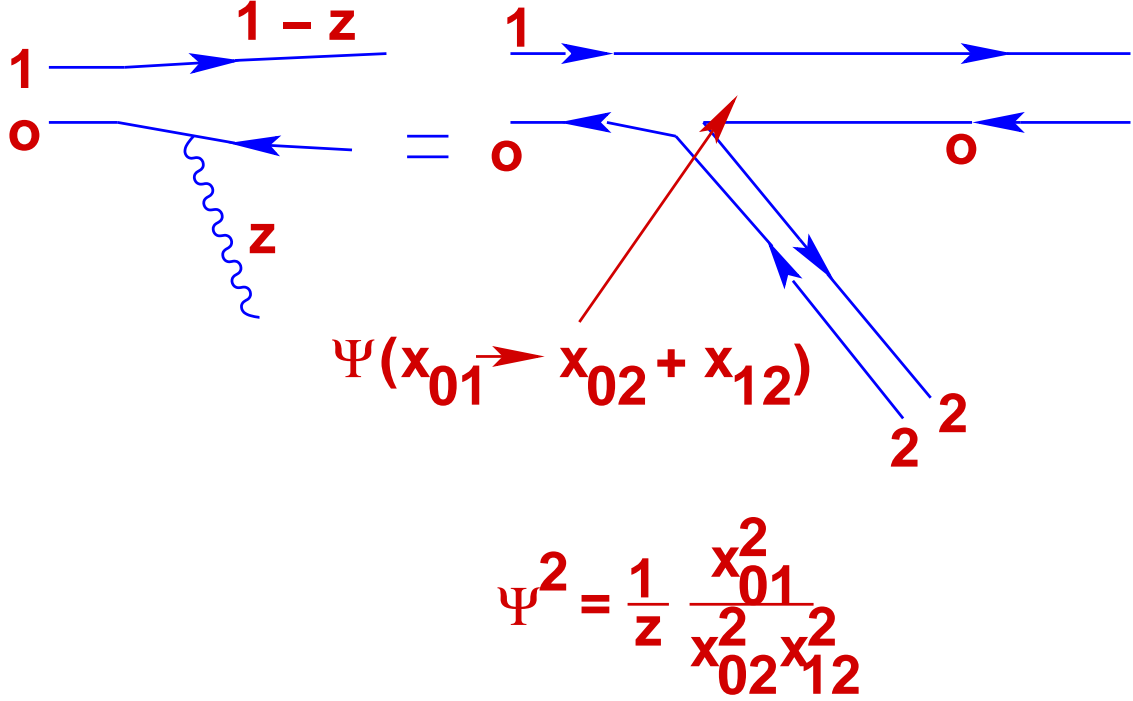}}
\end{center}
\caption{BFKL gluon emission as a  colour dipole decay.}
\label{fig:lev:dig}
\end{figure}

A. Mueller proved two results\cite{DOF3,MU94} which really showed that the
colour dipoles  are
the correct degrees of freedom in QCD at high  energies. First, he showed
that the gluon structure function can be viewed as the interaction of the
colour dipole with the target as shown in
Fig.~\ref{fig:lev:mgfor}. Secondly, he proved that the BFKL evolution can be
rewritten as a decay of one dipole into two dipoles for large $N_c$, as one can see in
Fig.~\ref{fig:lev:dig}.

Therefore, we can discuss the high energy (low $x$) DIS in terms of colour
dipoles which interact with themselves and with the target.

\subsection{Glauber-Mueller formula:}

It  turns  out that for an interaction  with the target we can obtain the simple
Glauber-Mueller formula which reads \cite{DOF1,DOF2,DOF3,AGL}\footnote{Giving
 credit to the authors of Refs. \cite{DOF1,DOF2,AGL} we refer to  this formula 
as the Glauber-Mueller formula because A. Mueller was the  first who  proved that the
gluon structure function can be described as rescatterings of a dipole. This
result changed the whole approach  to DIS by creating a transparent picture
of the 
interaction in QCD at high energies.}

\begin{equation}\label{eq:lev:gmf}
\sigma_{dipole}(x,r_{\perp}) = 2\,\int d^2 b_t \,\left(\,1\,\,-\,\,e^{- \frac{\Omega(x,r_{\perp};b_t)}{2}}\,\right)
\end{equation}
with opacity 
\begin{equation}\label{eq:lev:omega}
\Omega\,\,=\,\,\frac{\alpha_S(\frac{4}{r^2_{\perp}})}{3}\,\pi^2\,r^2_{\perp}\,\left(\,x\,
G^{DGLAP}(\frac{4}{r^2_{\perp}},x)\,\right)\,S(b_t)\,\,,
\end{equation}
where $S(b_t)$ is the target profile function. In the case of a  nucleon
target we can use the Gaussian form of $S(b_t) = (1/\pi R^2)\,e^{ -
  b^2_t/R^2}$ while for nuclei we use the Wood-Saxon parameterization
  for $S_A(b_t)$.

Eq. ~\ref{eq:lev:gmf} is a solution to the $s$-channel unitarity constraint
for the dipole-target amplitude
\begin{equation}\label{eq:lev:sun}
2\,Im \,a_{dipole}(x,r_{\perp};b_t)\,\,=\,\,|a_{dipole}(x,r_{\perp};b_t)|^2\,\,+\,\,G_{in}(x,r_{\perp};b_t)
\end{equation}
The inelastic cross section is equal to
\begin{equation}\label{eq:lev:inx}
\sigma^{in}_{dipole} \,\,=\,\,\int \,d^2 \,b_t \,\left(\,1\,\,-\,\,e^{- \Omega(x,r_{\perp};b_t)}\,\right)\,\,.
\end{equation}
Opacity $\Omega$ describes the interaction of one parton shower with the
target as one can see in Fig. ~\ref{fig:lev:mgfor}.

As has been mentioned the real breakthrough  was the proof by A. Mueller
that the gluon structure function can be calculated  using a  similar
formula for the colour dipole rescatterings, namely \cite{DOF3}
\begin{equation}\label{eq:lev:mggstr} 
x\,G(x, Q^2) \,\,=\,\,\frac{8}{\pi^3}\,\int^1_x\,\frac{d
  r^2_{\perp}}{r^4_{\perp}}\,\int \,d^2b_t\,
\left(\,1\,\,-\,\,e^{- \frac{(9/4)\Omega(x,r_{\perp};b_t)}{2}}\,\right)\,\,.
\end{equation}

\subsection{Packing factor:}
Eqs.~\ref{eq:lev:gmf}, \ref{eq:lev:inx} and \ref{eq:lev:mggstr} allow us  to introduce a
packing factor for colour dipoles in the parton cascade. This factor is equal
to
\begin{equation}\label{eq:lev:kappa} 
\kappa\,\,=\,\,(9/4)\,\Omega(x,r_{\perp};b_t =0)\,\,=\,\,
\frac{3\alpha_S(\frac{4}{r^2_{\perp}})}{4}\,\frac{\pi^2\,r^2_{\perp}}{\pi R^2}\,\left(\,x\,
G^{DGLAP}(\frac{4}{r^2_{\perp}},x)\,\right)
\,\,.
\end{equation}

The physical meaning of $\kappa$ is very simple: $\kappa =
\sigma_{dipole}/\pi  R^2$ $ = \sigma_{dipole}(BA) xG(x,Q^2)/\pi R^2$ $ =
\sigma_{dipole}(BA) \cdot \rho$. Therefore, $\kappa$ is the size of the
parton ( or preferable  to say its typical cross section ) multiplied by the
density of gluons in the transverse plane.

\subsection{Observables:}

It should be stressed that all our observables can be calculated if we know
the dipole amplitude. Indeed, the main obsevables  such   as the total
photon-hadron cross section , the gluon density and single diffraction
production inclusive cross section,   have a very simple
relation  with the dipole amplitude, namely,
\begin{eqnarray}
\sigma(\gamma^* p ) &=& \int^1_0\,d z \,\int \,d^2 r_{\perp}
 |\Psi(z, r_{\perp}; Q^2)|^2 \,\,
\sigma_{dipole}(x_B,r^2_{\perp})\,\,; \label{eq:lev:O1}\\
x G(x,Q^2) &=& \frac{4}{\pi^3}\,\int^1_x\,\frac{d
  x'}{x'}\,\int^{\infty}_{4/Q^2}\,\,\frac{d r^2_{\perp}}{ r^4_{\perp}}
  \,\,\sigma_{dipole}(x',2\,r^2_{\perp})\,\,;\label{eq:lev:O2}\\
\sigma^{SD}(\gamma^* p )&=&\int^1_0\,d z \int \,d^2 r_{\perp}
 |\Psi(z, r_{\perp}; Q^2)|^2 \int \,d^2 b_t | a_{dipole}(x,r_{\perp};b_t)|^2\,.\label{eq:lev:O3}
\end{eqnarray}
 
Eq.~\ref{eq:lev:O1} was proven in Refs.\cite{DOF1,DOF2,DOF3,AGL}, while  
Eq.~\ref{eq:lev:O2} was first written  in Ref.\cite{DOF3} and was discussed
in  detail in Ref.\cite{AGL}.   The  formula for the total diffractive
 production in
DIS was suggested in Ref.\cite{KM}.

The argument $2 r^2_{\perp}$ in Eq.~\ref{eq:lev:O2} reflects the fact that
actually the rescattering of gluon corresponds to rescatterings of two
dipoles of the same size,  which works effectively as the interaction of one
dipole but with a size which is $\sqrt{2}$  larger  (for $N_c \gg 1$). $\Psi$ was
calculated in Refs.\cite{DOF3,NZ}.
                                              
\section{Non-linear Evolution}

\subsection{The equation:}

\begin{figure}[h]
\begin{center}
  \epsfxsize=16cm
\leavevmode
\hbox{ \epsffile{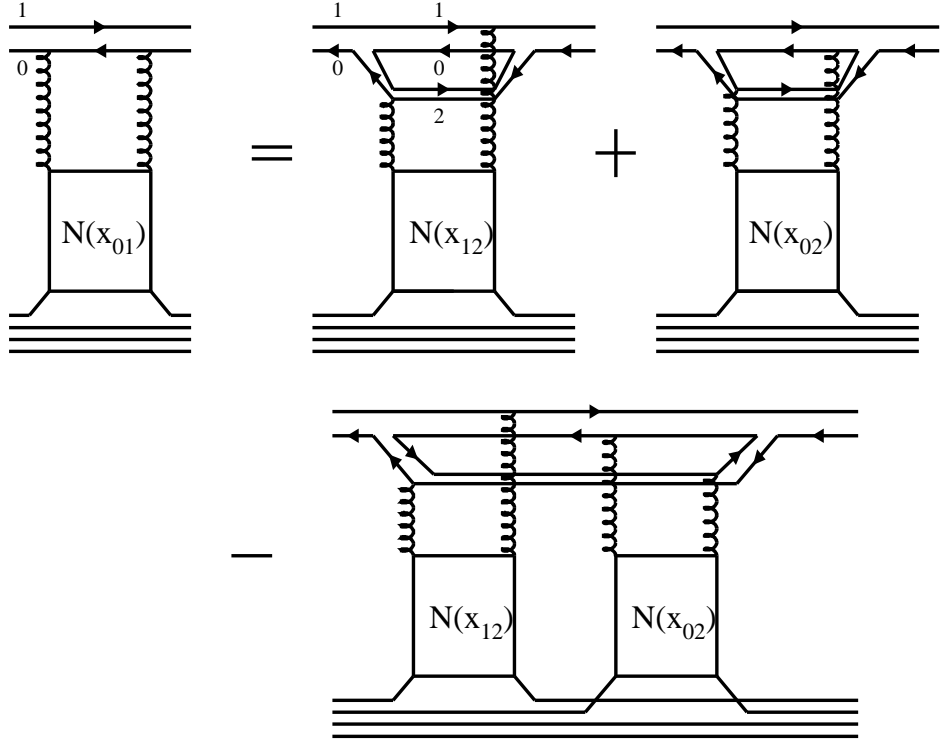}}
\end{center}
\caption{The pictorial form of non-linear evolution equation.}
\label{fig:lev:eq}
\end{figure}

Using the colour dipole picture of an interaction and, in particular, the fact
that the BFKL emission can be viewed as a decay of one colour dipole into
two  ( see Fig.~\ref{fig:lev:dig} ) we can easily obtain the nonlinear
equation for the imaginary part of the elastic dipole-target amplitude
$N(x,r_{\perp};b_t) = Im\, a^{el}_{dipole}(x,r_{\perp};b_t)$. This equation can
be written in the form (see also Fig.~\ref{fig:lev:eq}:
\begin{eqnarray} \label{eq:lev:glrint}
\frac{d N({\mathbf{x_{01}}},b_t,y)}{d y}\,\,\,&=&\,\,\,- \,\frac{2
\,C_F\,\alpha_S}{\pi} \,\ln\left(
\frac{{\mathbf{x^2_{01}}}}{\rho^2}\right)\,\,
N({\mathbf{x}},b_t,y)\,\,\,+
\,\,\,\frac{C_F\,\alpha_S}{\pi}\,\,
\int_{\rho} \,\,d^2 {\mathbf{x_{2}}}\,
\frac{{\mathbf{x^2_{01}}}}{{\mathbf{x^2_{02}}}\,
{\mathbf{x^2_{12}}}}\, \nonumber \\
 &\cdot&\,\,\,
\left(\,\,2\,N({\mathbf{x_{02}}},b_t,y)\, 
-\,N({\mathbf{x_{02}}}, b_t,y)\,N({\mathbf{x_{12}}},
 b_t,y)\,\right)\,\,, 
\end{eqnarray}
 where $y = \ln(1/x)$ and we assume that that $b_t \,\leq\, x_{02}$ or/and $x_{12}$.

Eq.~\ref{eq:lev:glrint} has a very simple meaning and actually describes the
fact that the dipole of the size $x_{01}$ decays into two dipoles of 
sizes $x_{02}$ and $x_{12}$ with probability $|\Psi|^2 =
\frac{x^2_{01}}{x^2_{02}\,x^2_{12}}$
as it is shown in Fig. ~\ref{fig:lev:dig}. These two dipoles then interact
with the target. They can interact separately and this interaction leads to
the linear term in Eq.~\ref{eq:lev:glrint}. However, two produced dipoles can
interact with the target simultaneously generating the non-linear term in the
equation. From Fig. ~\ref{fig:lev:eq} one can
see that this non-linear term takes into account the Glauber correction for
the 
two dipole interaction.The minus sign  in front of the non-linear term
reflects the  well known fact that we overestimate the value of cross section
considering it as a sum of two independent collision, since sometimes one
dipole happens to be in the shadow  of the second one.
 The linear term in  Eq.~\ref{eq:lev:glrint} is the
BFKL evolution,  which describes the evolution of the multiplicity of the fixed
size colour dipoles with respect to rapidity $y$. At first sight the linear
term sums the  leading twist contribution while the non-linear one is related to
higher twist contributions. However, this  is not true. The first term ( the
BFKL equation ) has also higher twist contributions but with the same
anomalous dimension as the leading twist ones. On the other hand, as  was
pointed out by Mueller and Qiu \cite{MUQI} the non-linear part
contributes mostly to the leading twist. The beauty of the equation is that it sums
both leading and higher twist contributions in a unique fashion and claims
that at any fixed $Q^2$ ( at any short distance ),  the higher twist
contribution will dominate  at sufficiently low $x$.

\subsection{Brief review of the theoretical approaches:} 

Eq.~\ref{eq:lev:glrint} shows that the problem of high density QCD has been
solved from  first principles and we think that it is instructive
to give a brief review of the theoretical approaches that all converge to
this equation.
\par
\begin{figure}[hptb]
\begin{tabular}{ c c}
\psfig{file=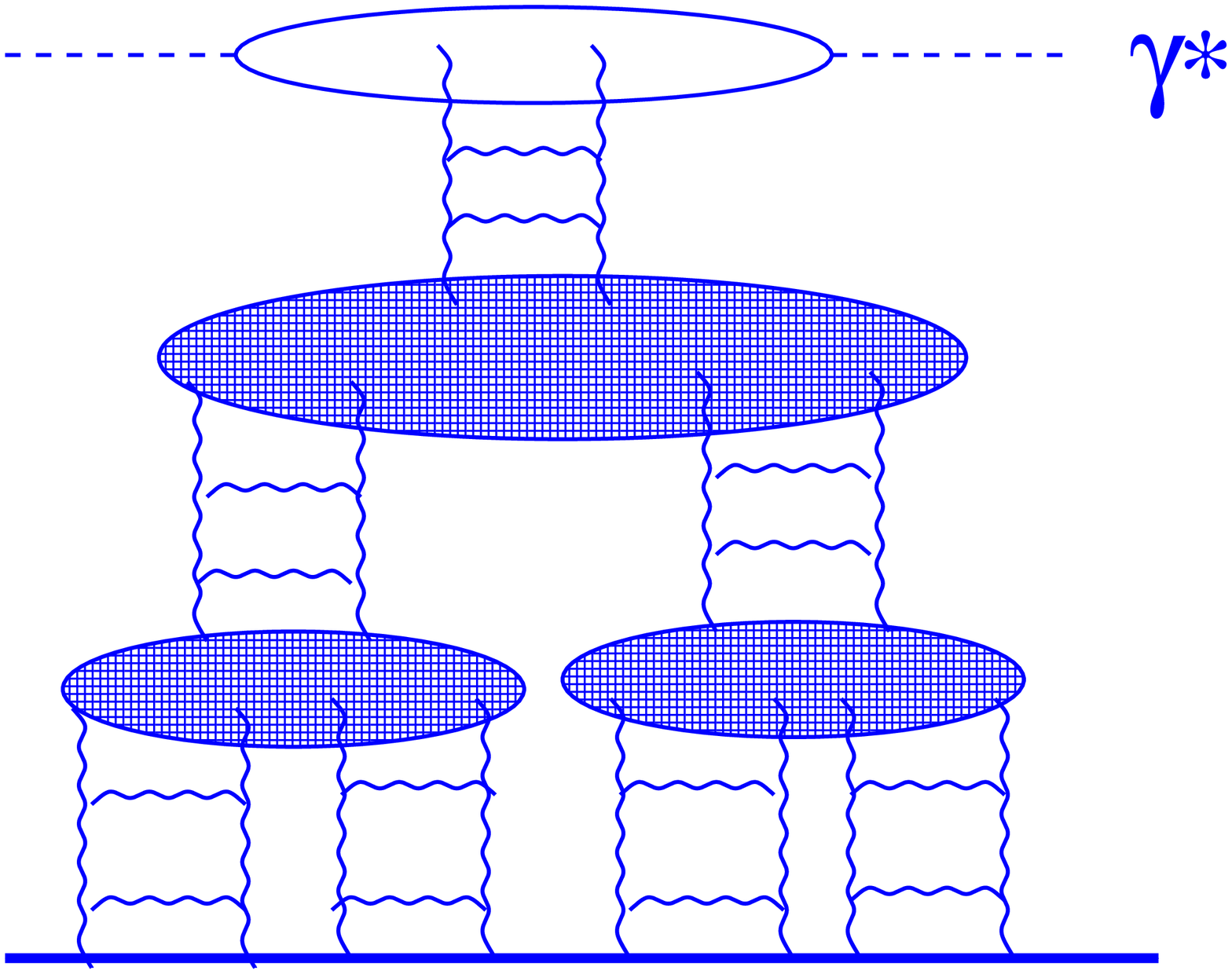,width=80mm} & \psfig{file=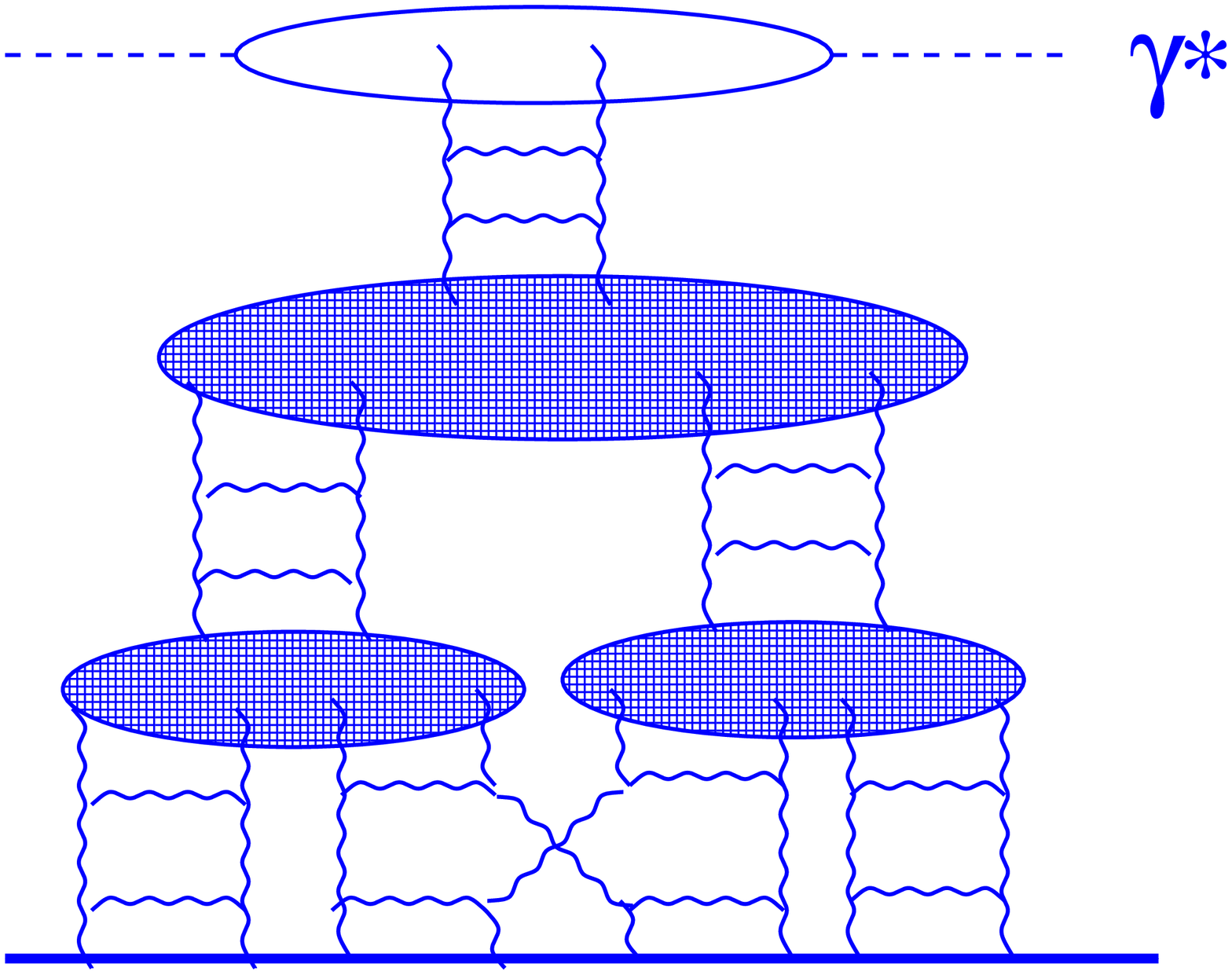,width=80mm}\\
       &  \\
a) &  b) \\
 &  \\
\end{tabular}
\caption{ ``Fan'' diagrams (a) and $1/N_c$ corrections to them (b).}
\label{fig:lev:fan}
\end{figure}

      
{\bf 1981 - 1983} \quad  GLR pointed out the new phase of QCD-high density QCD,
developed picture
  of  parton interaction in the Bjorken frame ( see above ), proposed
the  hypothesis  of parton saturation  and suggested first non-linear equation which
 sums
   the ``fan'' diagrams
  of Fig. ~\ref{fig:lev:fan}-a  and which  is actually
  Eq.~\ref{eq:lev:glrint}  in  momentum space\cite{GLR}.
\par 
{\bf 1986}  \quad  Mueller and Qiu \cite{MUQI} proved the GLR equation in the double log
approximation
 of perturbative QCD.

\par
{\bf 1992 - 1995} \quad J.Bartels\cite{BHT} showed that the non-linear equation
can be correct
  only in large $N_c$ approximation since $1/N_c$ corrections
(see  Fig. ~\ref{fig:lev:fan}-b ) 
 lead to the  interaction of two ladders in
the ``fan'' diagrams (see also Ref. \cite{LRHT}.
 Laenen and Levin based on Ref.\cite{LLS}  generalized the non-linear
equation, taking  into account  $1/N_c$ corrections in double log approximation
\cite{LL}.
 It turns out that $1/N_c$-approximation  works quite well in this problem and can be
treated  
using the generating function formalism.
\par
{\bf 1994} \quad  L. McLerran and R. Venugopalan \cite{MV} noticed that at high density
the gluonic  fields
 are strong (  $G_{\mu \nu} \approx \,1/g$ where $\alpha_s = g^2/4\pi$)
 and, therefore, one  can   approach 
 the high density QCD using the semiclassical gluon fields. Based 
on the
space-time structure
of the parton cascade  at low $x$,   they built the
effective Lagrangian for high density QCD.

\begin{figure}[h]
\begin{center}
  \epsfxsize=17cm
\epsfysize=8cm
\leavevmode
\hbox{ \epsffile{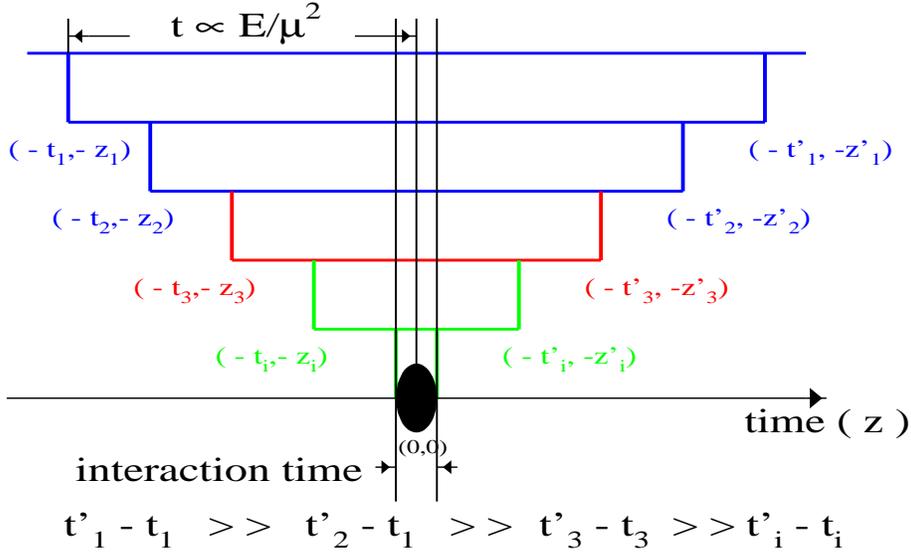}}
\end{center}
\caption{The space-time structure of the parton cascade at low $x$.}
\label{fig:lev:ststr}
\end{figure}

Indeed, the main and very important feature of the parton cascade, shown
in Fig.~\ref{fig:lev:ststr},
 is the fact that a parton with higher energy in the cascade lives  much
longer than a parton  with lower energy. We  follow   the parton emitted
at time $t'_2$. This parton
 lives a  much shorter  time than all partons emitted
before. Therefore, 
 all these partons  will have enough time to form a current
which depends  on their density.  Since the density is large we expect that the
current is a classical current. Finally, the Lagrangian of the interaction
for the parton emitted at time $t'_2$ can be written as  
$$ L(\rho)\,\,+\,\,j_{\mu}\cdot A_{\mu}\,\,+\,\,L(A)\,\,, $$
where $A$ is the field of a parton emitted at $t'_2$. However, we can
consider  a parton emitted at $t = t'_3$ and include the previous one in the
system with   density $\rho$. The form of the Lagrangian should be the
same. This is  a strong condition (equation) on the form of the effective
Lagrangian, so  called Wilson renormalization group approach. This is the 
beautiful idea of   the McLerran and  Venugopalan which leads to  
Eq.~\ref{eq:lev:glrint},     as we will  show  below.
\par
{\bf 1996} \quad I. Balitsky \cite{BA} proved the non-linear equation in Wilson Loop
Operator Expansion at high energies developed  by him. Unfortunately, his
paper was not noticed by the experts in the field including me. It
should be stressed that  he also  gave  an operator proof of the BFKL equation.
\par
{\bf 1997} \quad  Ayla, Ducati and Levin suggested non-linear equation
\cite{AGL} which     differs from that of
Eq.~\ref{eq:lev:glrint}. They used  the double log  approximation but summed
two DLA contributions

 ($( \alpha_s \ln(1/x) \ln( Q^2/\Lambda^2))^n$  and
$( \alpha_s \ln(1/x) \ln( Q^2_s(x)/Q^2))^n$.
It turns out that their equation
is just the same equation as Eq.~\ref{eq:lev:glrint}  but written for the
opacity $\Omega$  instead of $N = ( 1 - exp(- \frac{\Omega}{2}) )$. AGL
determined 
all numerical coefficients and discussed the initial condition of 
Eq.~\ref{eq:lev:glrint} which we will consider later.
\par
{\bf 1999 -2000} \quad  Yu. Kovchegov\cite{KOV} proved Eq.~\ref{eq:lev:glrint} in the
colour dipole approach \cite{MU94}.  It should be stressed that he not
only gave the derivation which we have discussed,  but he found a correct
observable which enters the equation ($N$) and he suggested an initial
condition that we will discuss below.
\par
{\bf 2000} \quad M. Braun \cite{BR} calculated the ``fan'' diagrams of
Fig.~\ref{fig:lev:fan}-a in the BFKl  kinematic region, using the triple
ladder vertex of Refs. \cite{3P}.
\par
{\bf 2001} \quad 
Iancu,
 A. Leonidov and L. McLerran \cite{ILM} (see also
Ref. \cite{IM}) proved   Eq.~\ref{eq:lev:glrint} in the effective
Lagrangian approach. Their proof is  based on  long and successive 
development of the effective Lagrangian approach exploited  in
Refs. \cite{EFLA}

\subsection{Initial conditions:}

One can see that Eq.~\ref{eq:lev:glrint} does not depend on the target and
the  dependence on the target comes from the initial conditions at some initial
value of $x = x_0$. For a target nucleus it was argued \cite{AGL,KOV} that
the initial conditions should be taken in the  Glauber - Mueller form (see
Eq.~\ref{eq:lev:gmf}), namely,
\begin{equation} \label{eq:lev:incon}
N(x_{01},x=x_0,b_t) \,\,=\,\,N^{GM}(x_{01},x=x_0,b_t) =
1\,\,-\,\,e^{-\,\frac{\Omega(x_{01},x=x_0,b_t)}{2}}\,\,.
\end{equation}
The value of $x_0$ is chosen in the interval 
$$ e^{- \frac{1}{\alpha_S}}\,\leq\,x_0\,\leq\,\frac{1}{2\, m\,R}
\,\,,$$
where  $R$ is the radius of the target.  In this region the value of $x_0$ is
small enough to use the low $x$ approximation, but the production of the gluons
(color dipoles) is still suppressed as $\alpha_S \ln (1/x)
\,\leq\,\,1$.
Therefore, in this region we have the instantaneous exchange of the
classical gluon fields.
Hence, an incoming
color dipole interacts separately with each  nucleon in a nucleus (see
Ref. \cite{MUKO}).

For the hadron we have no proof that Glauber-Mueller formula is correct.  
 As far
as we understand the only criteria in this problem (at the moment) is the
correct
description of the experimental data. We described \cite{GLMAM}     almost all available
 HERA data using Eq.~\ref{eq:lev:gmf},  and we feel confident using
 Eq.~\ref{eq:lev:gmf} as the  
initial
condition for Eq.~\ref{eq:lev:glrint}. It should be stressed that the
experimental data on $d F_2/d \ln Q^2$ provides   direct information on the
integral over $b_t$ for $N$ since \cite{AGLFRT}
\begin{equation} \label{eq:lev:slp}
\frac{d \,F_2(x,Q^2)}{d \ln Q^2}\,\,=\,\,\frac{Q^2}{3 \pi^3}\,\,\int \,d^2
\,b_t\, N(x,4/Q^2,b_t)\,\,+\,\,O(1/ln Q^2)\,\,.
\end{equation}
Choosing $x_0 = 10^{-2}$ we can make the initial condition  practically
independent of the parameterization of the $xG^{DGLAP}$ since all available
parameterizations give the same prediction,  even for the gluon density in
this
$x$ - range. 
\subsection{Theory status of the approach:}
\subsubsection{Parameters of the approach:}

The master equation (Eq.~\ref{eq:lev:glrint}) as well as the Glauber -
Mueller formula (see  Eq.~\ref{eq:lev:gmf} and Fig. \ref{fig:lev:mgfor} )
sums all diagrams of the
 order of
\begin{equation} \label{eq:lev:ts1}
\left(\alpha^2_S \,(\frac{1}{x}
)^{\Delta}\,\right)^n\,\,\,{\rm with}\,\,\,\Delta \,\,\propto\,\,\alpha_S
\end{equation}
It means that starting from $\alpha_S\ln (1/x)
\,\,\approx\,\,\,\ln(1/\alpha_S)$ \footnote{ This estimate comes from
  $\alpha^2_S \,(\frac{1}{x})^{\Delta}\,\,\approx\,\,1$.}
the corrections related to the rescatterings of the partons and their
interactions become essential. On the other hand, the next to leading order
corrections to  DGLAP or  BFKL equations produce a  value of $\Delta$ in
Eq.~\ref{eq:lev:ts1} calculated up to $\alpha^2_S$ accuracy (   $\Delta = C_1 \alpha_S
+ C_2 \alpha^2_S$ ). Such corrections start to become important only for
$\alpha_S\ln (1/x) \,\geq \,\,1/\alpha_S$ or , in other words, the next to
leading order corrections have to be calculated  after the
rescatterings and parton interactions, included in the master equation, have
been taken into account. 

Therefore, {\bf the calculations of the next to leading order corrections
for the linear evolution equations can be considered as a QCD motivated model}
in which  crucial {\it ad hoc } assumptions have been made, namely, that the parton
interactions and their rescatterings are small numerically. 

\subsubsection{Large $\mathbf{N_c}$ approach:}

\begin{figure}[h]
\begin{center}
  \epsfxsize=17cm
\epsfysize=4cm
\leavevmode
\hbox{ \epsffile{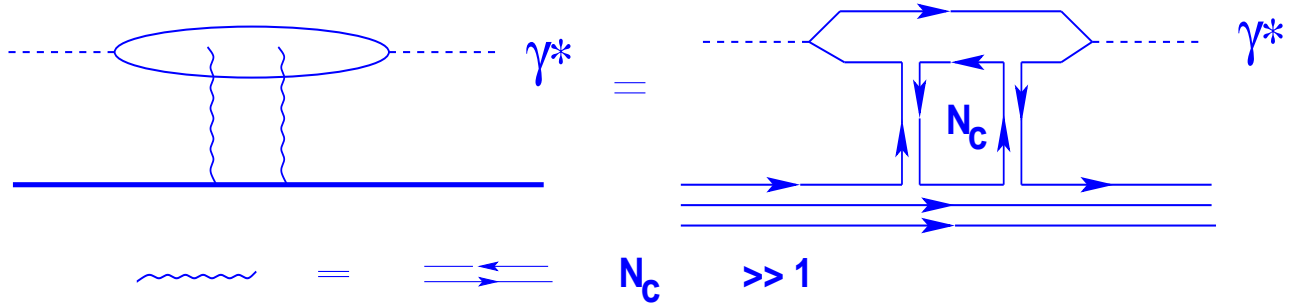}}
\end{center}
\caption{The Borm approximation for the interaction of the virtual photon
  with the proton in large $N_c$ limit of perturbative QCD.}
\label{fig:lev:lnc}
\end{figure}

As has been discussed the 
 master equation (Eq.~\ref{eq:lev:glrint}) has only been proved  in the  large
 $N_c$ limit of QCD. The key question arises: could we develop a
 selfconsistent large $N_c$ approach for the DIS at low $x$. For  $N_c
 \,\gg\,1$ we assume that $ \hat{\alpha}_S \,\equiv \,\alpha_S \,N_C \,\,\approx\,1$ while $\alpha_S
 \,\ll\,1$,  we assume all contributions that are  proportional to
 $\alpha_S$ to be
 small so that we can  neglect them. At first sight we have a problem with developing
 such an approach for high energy scattering. Indeed, the first Born diagram (
 see Fig.  \ref{fig:lev:lnc} ) turns out to be of the order of $(N_f/N_c)
 \hat{\alpha}^2_S$ \footnote{In large $N_c$ limit we can look at gluon as the
   quark-antiquark pair and Fig.  \ref{fig:lev:lnc} shows only one quark
   loop, which gives factor $N_c$, for the diagram which is of the order of $
   \alpha^2_S$.} where $N_f$ is the number of the quarks inside the proton.
 We can consider this parameter  as being  small  in the
 large $N_c$ limit. However, it is well known that a nucleon consists of the
 $N_c$ ( $N_f = N_c$) number of quarks. This observation makes 
  $(N_f/N_c)\hat{\alpha}^2_S$ a `small parameter' of the order of $
\hat{\alpha}^2_S$.

The amplitude $ N({\mathbf{x_{01}}},b_t=0,y) \,\,\approx
\,\,\hat{\alpha}^2_S/\pi R^2$ where $R$ is the radius of the proton for  $N_c
\,\gg\,1$.  For large $N_c$ the interaction of two quarks is proportional to
$\alpha_S$ and can be considered as small, but in the Hartree-Fock
approximation  it  leads to a potential
proportional to $\alpha_S \,N_f = \alpha_S \,N_c$.
 Therefore, $R \propto \hat{\alpha}_S$. 

Finally, we conclude that the master equation sums all contributions of the
order $ \left(\hat{\alpha}_S\,\ln(1/x) \right)^n$ and gives a selfconsistent
large $N_c$ limit for the DIS with the nucleons. In such an  approach the DIS
with mesons is suppressed by extra factor $1/N_c$.  However, the equation
itself depends only on $\hat{\alpha}_S$,  and $1/N_c$
suppression for the mesons means that the initial condition could be written
only as the first term of the Glauber-Mueller formula (see 
Eq.~\ref{eq:lev:gmf} ). Using this initial condition, we  find the first
$1/N_c$ corrections to the amplitude which will give the main contribution
for the high energy ( low $x$ ) asymptotic.

 It should be stressed that we
still do not need to take  $1/N_c$ corrections related to more
complicated diagrams of the Fig.~\ref{fig:lev:fan}-b type into account. We would like to
emphasize that the regular procedure, of  how to calculate $1/N_c$ corrections, has
been developed \cite{LL,BA,IM} and the first estimates have been made \cite{LL}.

\subsubsection{The region of applicability:}

To evaluate the value of energy up to which we can use the master equation as
a good theoretical tool for finding the high energy asymptotic,  we need to
estimate the contributions that have been neglected. We  discuss
all of them of the order of $1/N_c$, and they can be divided in two major
classes:

\begin{itemize}
\item \quad The first one is the interaction between two ``ladders'' of
type   Fig.~\ref{fig:lev:fan}-b . They are proportional to $\alpha_S/N^2_c
  \,\ln(1/x)$ \cite{BHT}, and therefore, we have a first restriction on the
  value of $x$:
\begin{equation} \label{eq:lev:ts2}
\frac{\hat{\alpha}_S}{N^2_c} \,\ln(1/x)\,\,<\,\,1
\,\,\,{\rm or}\,\,\,\ln(1/x)\,\,<\,\,\frac{N^2_c}{\hat{\alpha}_S}\,\,;
\end{equation}
\item \quad The second constraint comes from so called enhanced diagrams (see
  for an example 
  Fig.~\ref{fig:lev:enhance}  ).\, The ratio of the contribution of   this diagram  to the ``fan'' diagram of
Fig.~\ref{fig:lev:fan}-a is
 proportional to $\hat{\alpha}_S\,/N^2_c\,\ln(1/x)$ which leads to the same
 value at the highest energy,  as does
 Eq.~\ref{eq:lev:ts2}.

 \begin{figure}[h]
\begin{center}
  \epsfxsize=8cm
\leavevmode
\hbox{ \epsffile{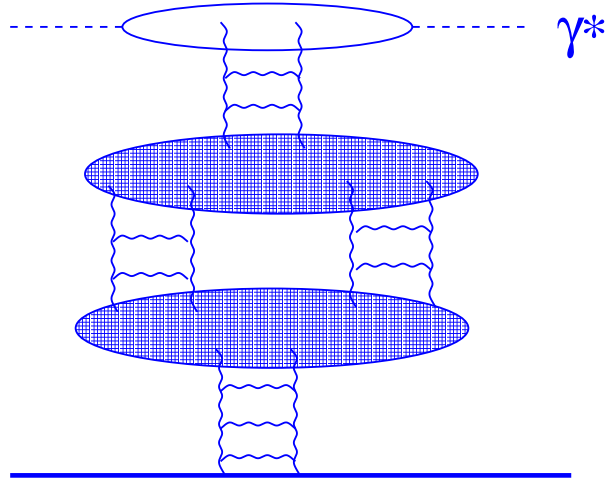}}
\end{center}
\caption{The first enhanced diagram in the parton cascade.}
\label{fig:lev:enhance}
\end{figure}

\end{itemize}
 Eq.~\ref{eq:lev:ts2} gives a sufficienly large value of energy upto which we
 can trust the master equation. The actual situation is much better. Indeed, the
 nonlinear equation leads to $N({\mathbf{x_{01}}},b_t,y)
 \,\longrightarrow\,1$ at high energy, and we will argue below that this
 asymptotic limit will be reached at much lower energy than one defined by  Eq.~\ref{eq:lev:ts2}.
Since $N({\mathbf{x_{01}}},b_t,y) = 1$ is the
 unitarity boundary for the scattering amplitude, the $1/N_c$ corrections
 cannot modify this result but could lead to a slightly different value of the
 saturation scale. 

It should be stressed that the regular procedure of  how to include $1/N_c$
corrections to the master equation,  has  been developed \cite{LL,BA,ILM,IM}
and even first estimates of the quantative effect of the $1/N_c$ corrections
have been performed \cite{LL}.

\section{Saturation scale $\mathbf{Q_s(x)}$}
\subsection{Simple estimates:}
\begin{figure}[h]
\begin{minipage}{12cm}
\begin{center}
\epsfxsize=10.4cm
\leavevmode
\hbox{\epsffile{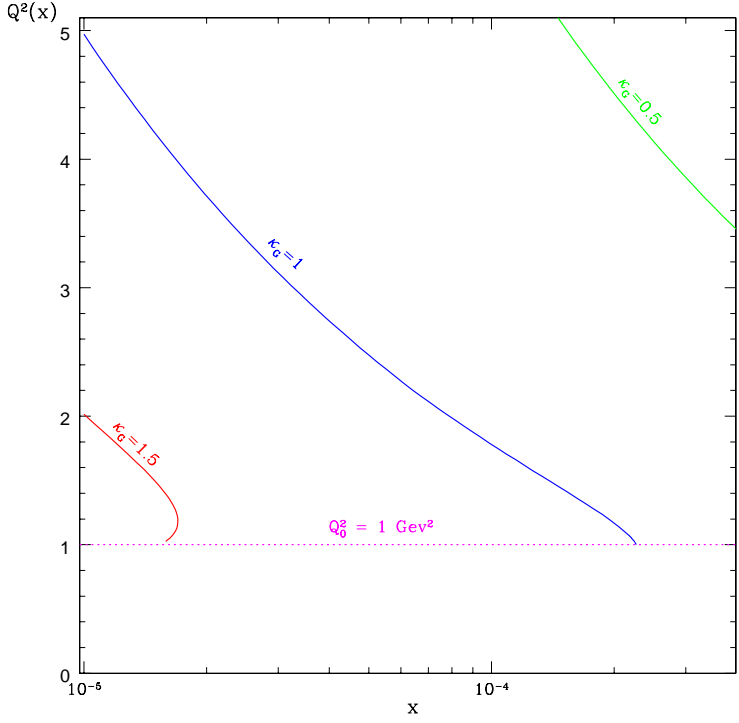}}
\end{center}
\end{minipage}
\begin{minipage}{4.5cm}
\caption{The  simplest
estimate for the saturation scale $Q_s(x)$ from
  equation $\kappa(x,Q^2_s(x))\,=\,1,\,\,1.5\,\,0.5$.}
\label{fig:lev:Qs}
\end{minipage}  
\end{figure}

The simplest estimates for the saturation scale come from the equation that
the packing factor $\kappa$ is of the order of 1 \cite{DOF3,AGL,GLMAM}.  
Fig. \ref{fig:lev:Qs}
shows  the solution of the equation
\begin{equation}\label{eq:lev:Qs}
\kappa(x,Q^2_s(x))\,=\,1
\end{equation}
As one can see  $Q_s$ can be rather large even in HERA region. However, this
estimate depends  crucially  on the value of $xG^{DGLAP}$.

\subsection{Analytic solution of the non-linear equation:}
\begin{figure}[h]
\begin{tabular}{c c}
\psfig{file=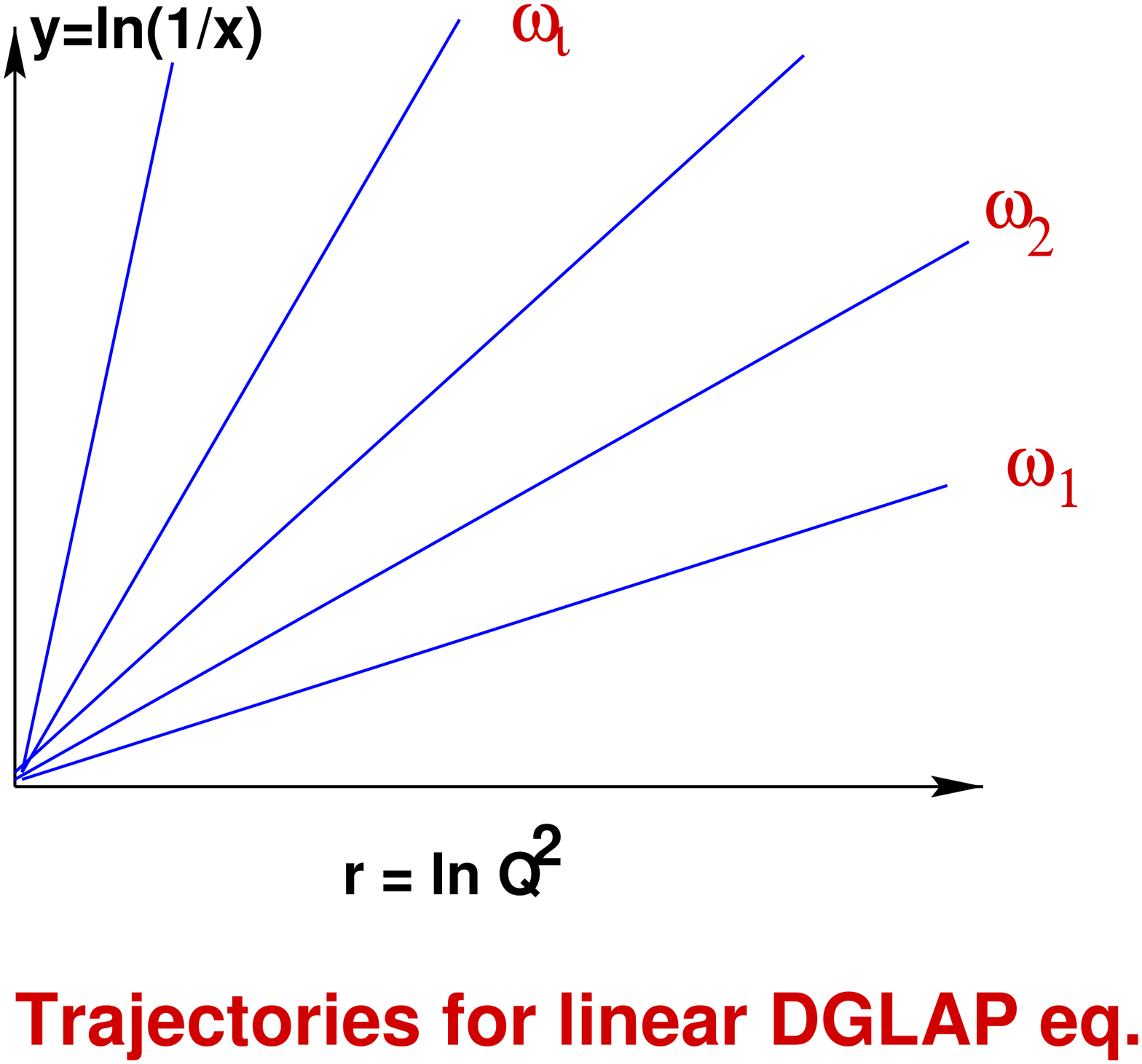,width=80mm,height=60mm} &
\psfig{file=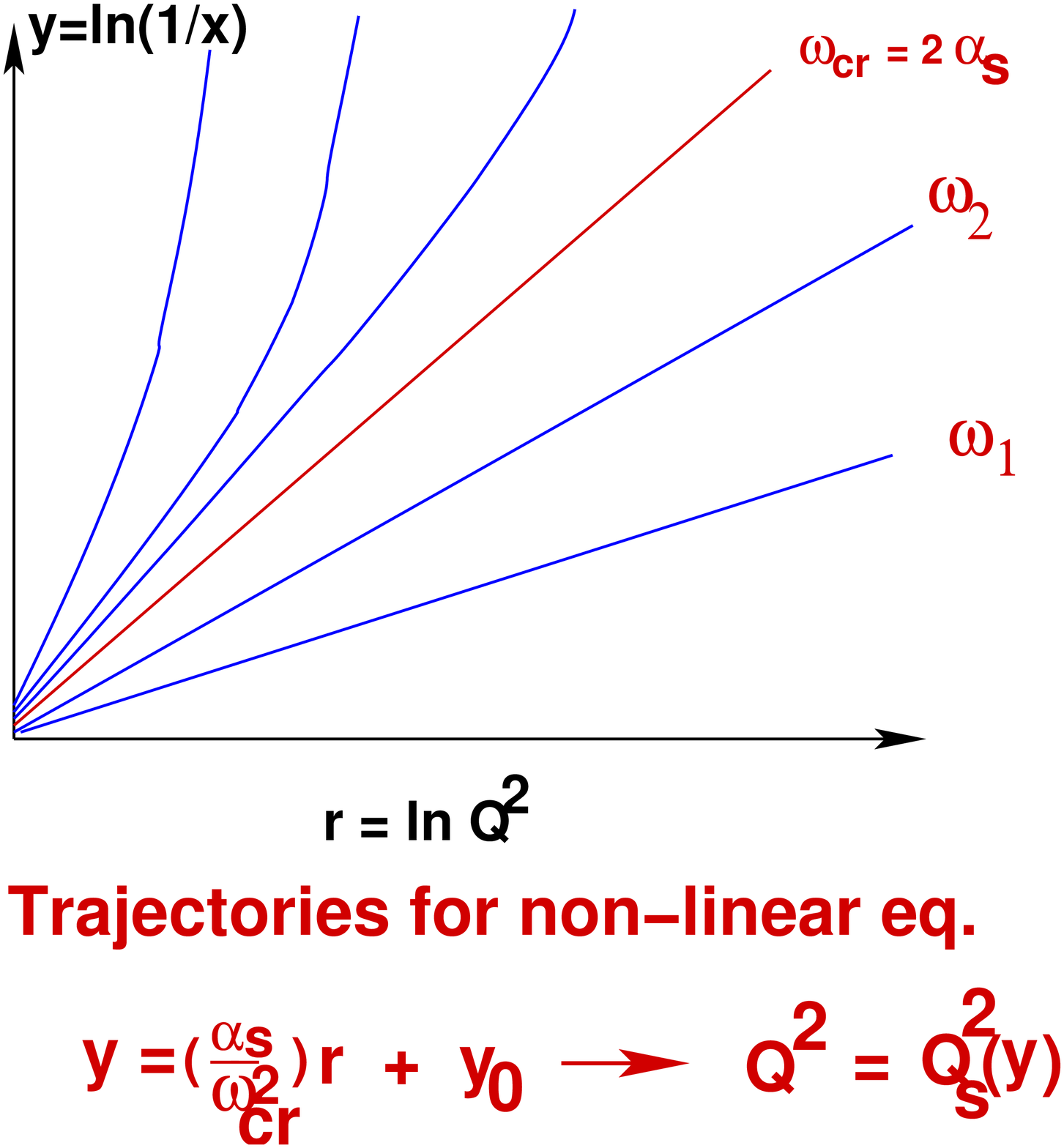,width=80mm, height=60mm}\\
       &  \\
a) &  b) \\
 &  \\
\end{tabular}
\caption{ Trajectories  for linear (a) and non-linear (b) equations.}
\label{fig:lev:traj}
\end{figure}

 Eq.~\ref{eq:lev:glrint} has been solved analytically with the  simplified version
 of the BFKL kernel \cite{LT,KOV1,IM}, namely, the  Mellin transform $\chi(f)$
 was
 taken as $\frac{1}{f} + \frac{1}{1 -f}$ instead of $\chi(f) = 2 \psi(1) -
 \psi(f) - \psi(1 -f)$,  where $\psi $ is the derivative of the log of Euler
 gamma function. In terms of $s$-channel resummation  Eq.~\ref{eq:lev:glrint}
 with this kernel sums two kinds  of double logs: $( \alpha_S
 \ln(1/x)\,\ln(Q^2/\Lambda^2) )^n$ for $Q^2 > Q^2_s(x)$ and $( \alpha_S \ln(1/x)\,\ln(Q^2_s(x)/Q^2) )^n$
for $Q^2 < Q^2_s(x)$. For large $Q^2$ Eq.~\ref{eq:lev:glrint} can be reduced
to the linear equation and can be solved using the trajectory
(characteristics ) method \cite{GLR,COLBAR}. The structure of the  trajectories
both for linear and nonlinear equation is shown in Fig.~\ref{fig:lev:traj}.

The idea is to interpret  the last linear  trajectory ( line with $\omega_{cr} =2\,\alpha_S$ 
  in Fig.~\ref{fig:lev:traj}-b ) of the linear equation which 
can be treated without nonlinear corrections,  as the saturation   scale. Doing
so, we obtain \cite{IM,LT} the following expression for the saturation scale
\begin{equation} \label{eq:lev:satsc}
Q^2_s(x)
\,\,=\,\,Q^2_s(x=x_0)\,\cdot\,\left(\,\frac{x_0}{x}\,\right)^{\frac{4 N_c \alpha_S}{\pi}}
\end{equation}
where $Q^2_s(x=x_0))$ is a saturation scale in our initial condition. 

Therefore, we expect power-like rise of the saturation scale at low $x$.

\subsection{Phenomenological saturation scale:}

Golec-Biernat and Wuesthoff \cite{GW} suggested one could  extract the value of
$Q_s(x)$ from HERA data assuming that the saturation region has been reached
at HERA. Surprisingly   they managed to describe  almost all HERA data using
a simple parameterization for the saturation scale, namely,
\begin{equation} \label{eq:lev:gwsat}
Q^2_s(x)\,\,=\,\,(1\,\,GeV^2)\,\cdot\,\left(\,\frac{x_0}{x}\,\right)^{\lambda}\,\,
\end{equation}
with $\lambda = 0.288$ and $x_0 = 3.04 \times 10^{-4}$.

\subsection{Saturation scale from numerical solution of the non-linear
  equation:}

\begin{figure}[h]
\begin{minipage}{12cm}
\begin{center}
\epsfxsize=10.4cm
\leavevmode
\hbox{\epsffile{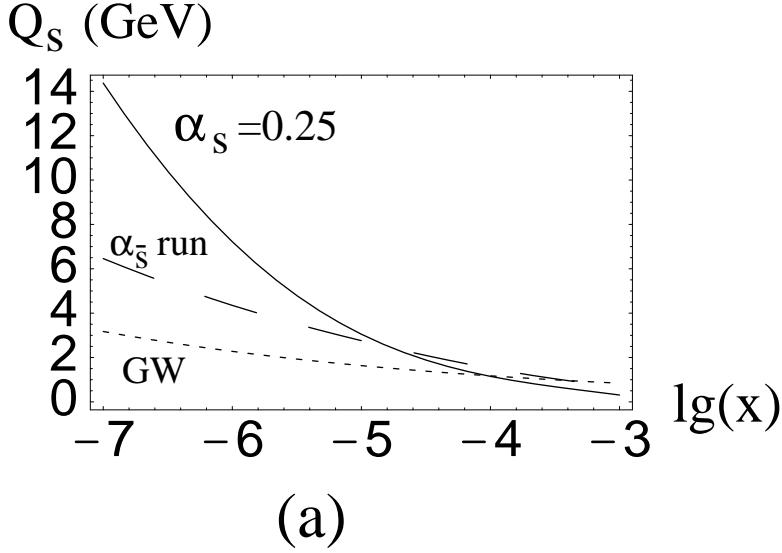}}
\end{center}
\end{minipage}
\begin{minipage}{4.5cm}
\caption{The   saturation scale $Q_s(x)$ plotted as a function of $lg x =
  log_{10}(x)$. GW denotes the saturation scale from Eq.~\ref{eq:lev:gwsat}}
\label{fig:lev:satnum}
\end{minipage}  
\end{figure}

In Ref. \cite{GLMNL} an  attempt was made to solve the non-linear equation
numerically (see also Ref. \cite{BR} ), starting from initial $x = x_0 =
10^{-2}$.
Defining  the saturation scale as a value of $r^2_{\perp} = 4/Q^2_s(x)$ at
which the imaginary part of the elastic amplitude for the dipole-target
scattering is equal to 1/2 
$$ N(r_{\perp}= 2/Q_s,x)= \frac{1}{2}$$
the saturation scale shown in Fig.~\ref{fig:lev:satnum} was calculated.

 One can see that  the saturation scale reaches a sufficiently  large value of
 the order of $Q_s(x) \approx 14\,GeV$ at $x = 10^{-7}$. It should be
 stressed that even at $x= 10^{-5}$ which is in the  typical range for THERA, the
 saturation  scale is approximately  2 $\div$ 3 times larger than the Golec-Biernat
 and Wuesthoff estimates.
\subsection{Saturation scale from  DIS with nuclei:}
\begin{figure}[h]
\begin{tabular}{c c}
\psfig{file= 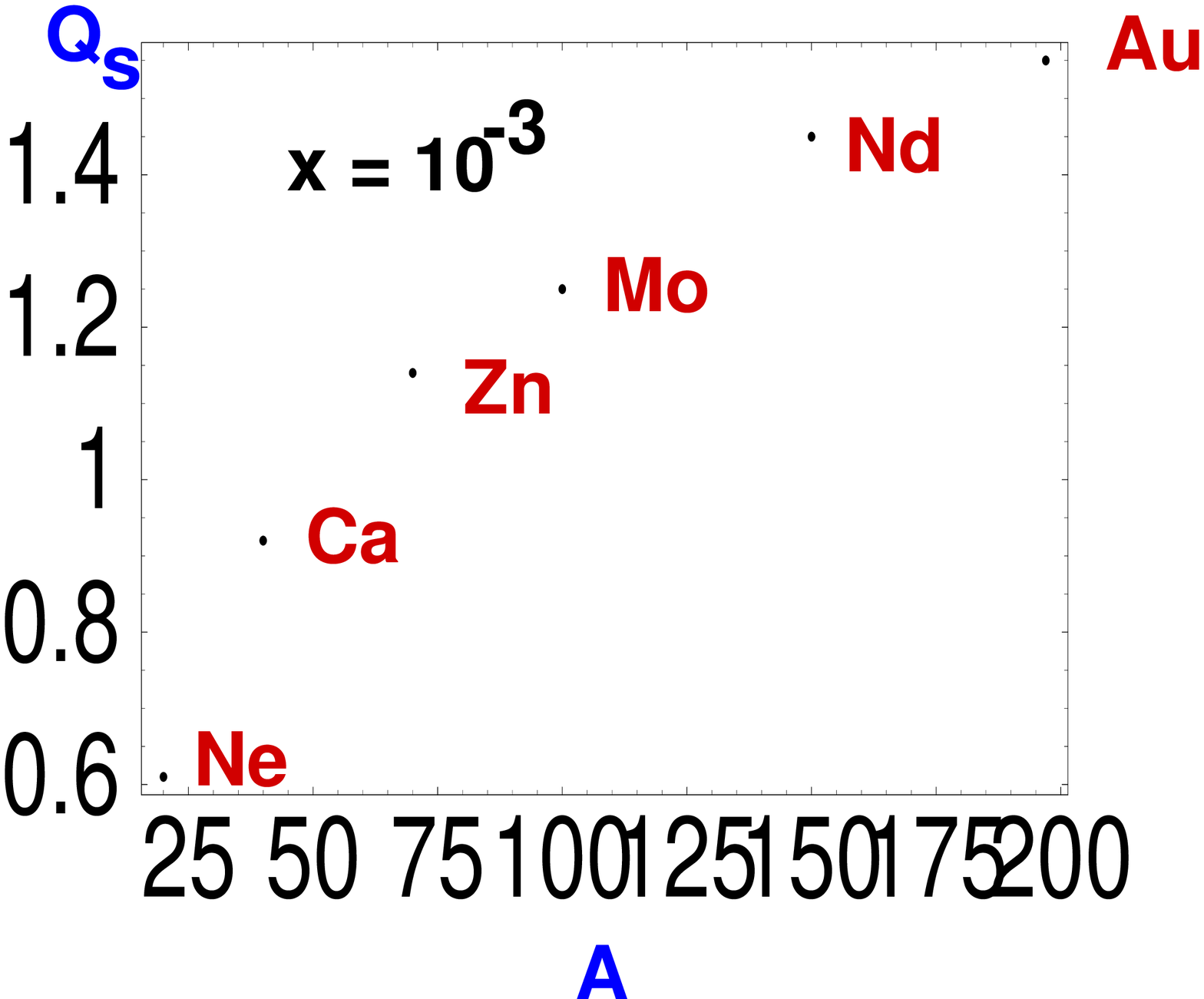,width=80mm,height=70mm} &
\psfig{file=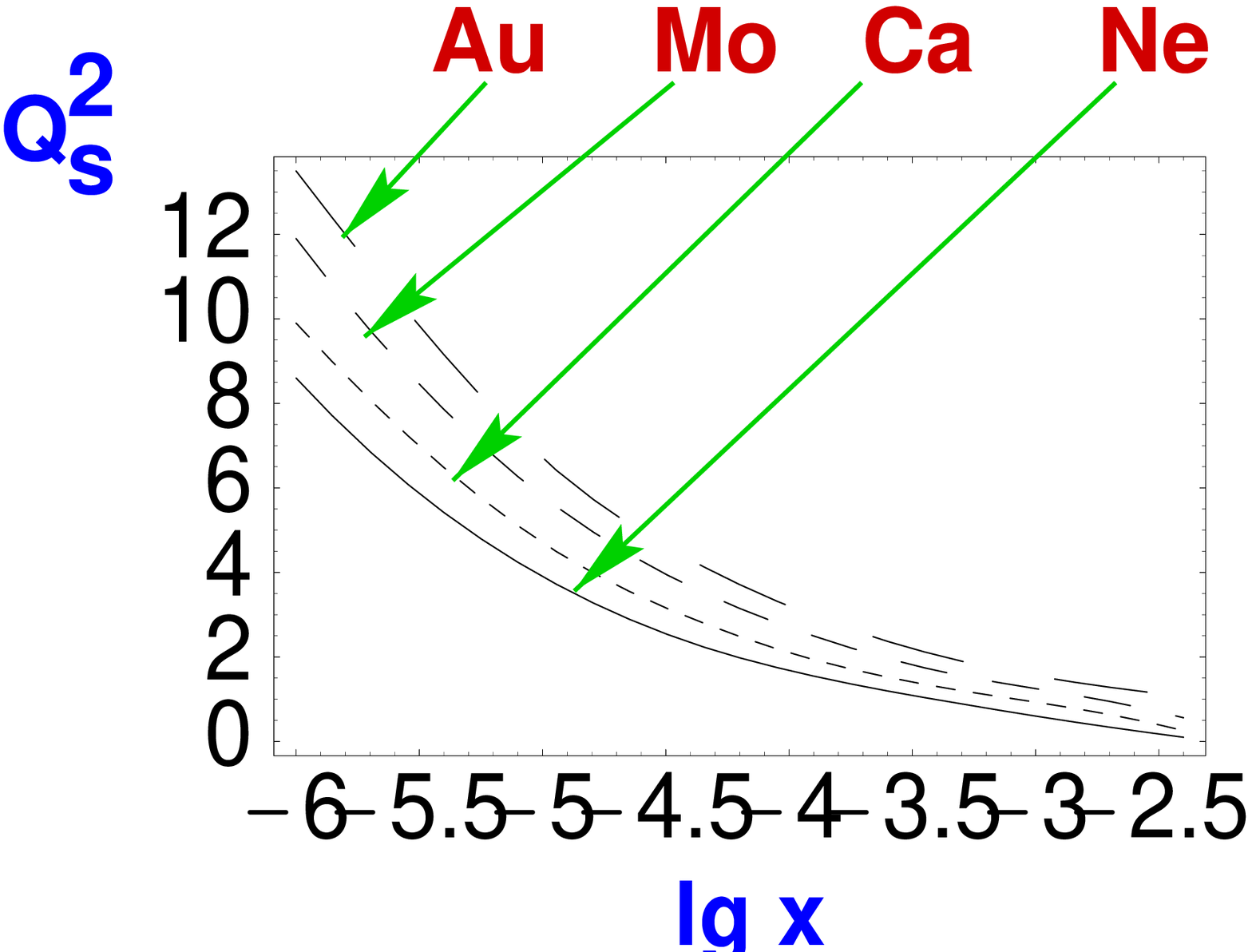,width=80mm, height=70mm}\\
       &  \\
a) &  b) \\
 &  \\
\end{tabular}
\caption{ Saturation scale for DIS with nuclei: A (a) and $x$ (b) dependencies .}
\label{fig:lev:satA}
\end{figure}

The first estimates for the saturation scale in DIS on  nuclei, presented in
Fig.~\ref{fig:lev:satA}, shows that the nucleus target gives   a promising
avenue to increase the parton density without requiring a    region of
extremely low $x$. Comparing Fig.~\ref{fig:lev:satA}
with Fig.~\ref{fig:lev:satnum} we  see that $Q_s(x)$ at $x=10^{-3}$ for gold
is almost twice  larger than for a nucleon. 

\section{A new scaling in the saturation region.}
\begin{figure}[h]
    \begin{center}
   \epsfxsize=15.5cm
      \leavevmode
    \hbox{\epsffile{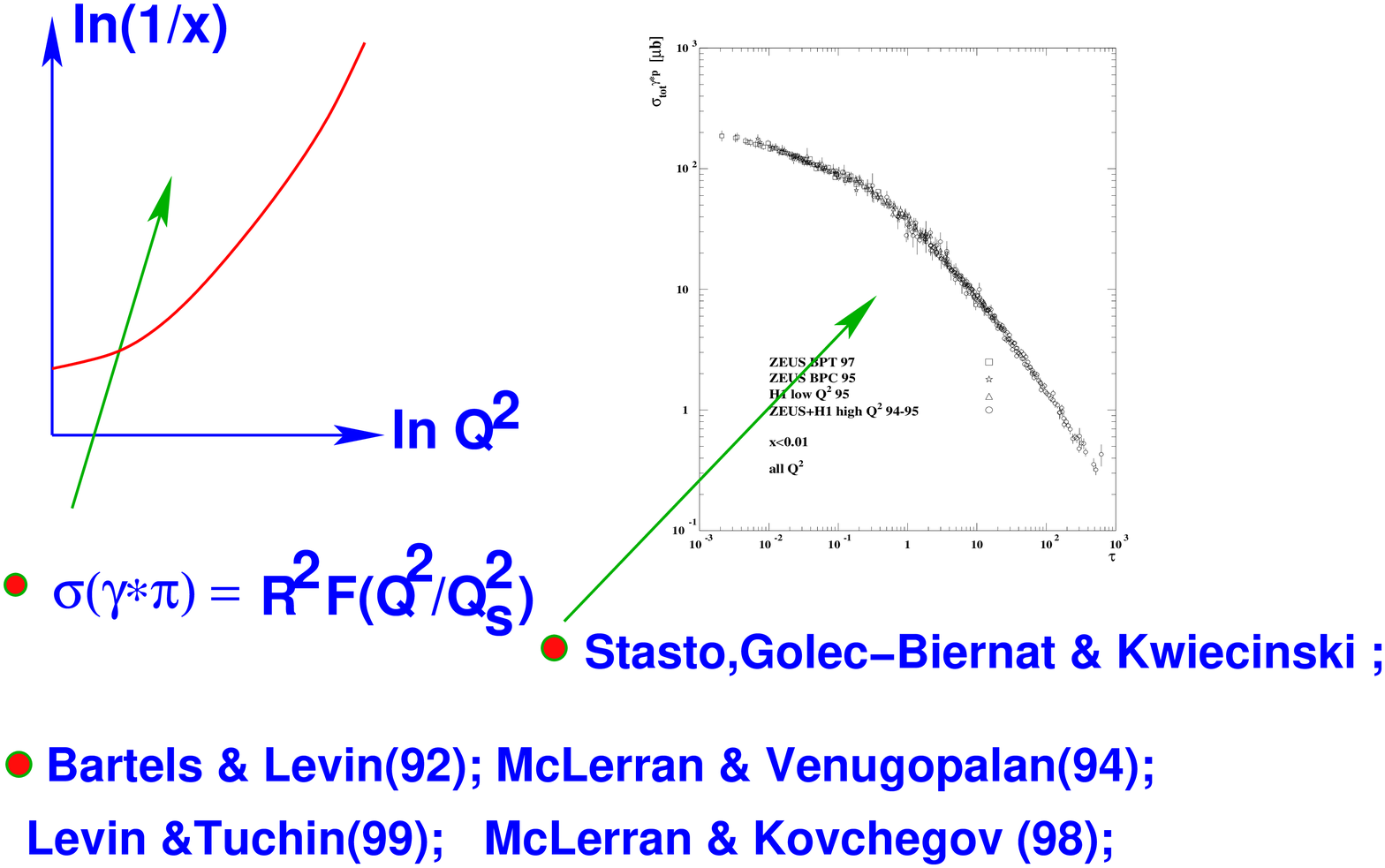}}  
    \end{center}
 \caption{}
      \label{fig:lev:nsc}
  \end{figure}

The simple picture of the  hadron  in the saturation region shown in Fig.~\ref{fig:lev:sat},
leads to a new scaling phenomenon    in this saturation region. We  expect that the
parton densities as well as cross sections  are not  functions of two
variables:$x$ and $Q^2$,  but they depend only on ratio $Q^2/Q^2_s(x)$
\cite{GLR,BL,MV,KM,LT}. Stasto, Golec-Biernat and Kwiecinski \cite{SGK} found
that this scaling is valid  for HERA data at $x < 0.01$. Fig.~\ref{fig:lev:nsc}
summarizes the situation and, here, I would like to comment on the
theoretical arguments for such a new scaling scheme.

\subsection{  Simple  arguments for a new scaling:}
 Let us rewrite Eq.~\ref{eq:lev:glrint} in momentum space where it looks simpler
\begin{equation} \label{eq:lev:momspglr}
\frac{d N(Q,y,b_t)}{d y} = \bar \alpha_S  \left( \int K(Q,Q') N(Q',y;b_t)
- N^2(Q,y;b_t) \right) \,\,.
\end{equation}
 For moments $N(\omega,Q)= N(\omega,Q_0) \,e^{\gamma(\omega)\ln Q^2}$
 Eq.~\ref{eq:lev:momspglr} can be reduced to the form
 \begin{equation} \label{eq:lev:omspglr}
\omega
e^{\gamma(\omega) ln Q^2} = \chi(\gamma) e^{\gamma(\omega) ln Q^2} -
N(\omega,Q_0) \int d \omega'  e^{ (\gamma(\omega') + \gamma(\omega - \omega')
 ln Q^2}               
\end{equation}
The integral over $ \omega'$ in Eq.~\ref{eq:lev:omspglr} can be evaluated  by
saddle point method 
$$
 \int d \omega'  e^{ (\gamma(\omega') + \gamma(\omega - \omega')
 ln Q^2}\,\,\,  \Longrightarrow\,\,\, e^{2\gamma(\omega/2) ln Q^2}
$$

Therefore, we can see two different regions in the solution to
Eq.~\ref{eq:lev:momspglr}: large $Q^2$, where the non-linear term is small and
can be neglected,  and low $Q^2$  where linear and nonlinear terms should be of
the same order. It gives \cite{BL}
\begin{equation} \label{eq:lev:gamat}
\gamma(\omega) = 2 \gamma(\omega/2) \,\,\quad\,\, \rightarrow\,\,\,
 \gamma(\omega)  = C \omega 
\end{equation}

Using Eq.~\ref{eq:lev:gamat} we obtain for  $ N(y,Q^2)$ 
\begin{equation} \label{eq:lev:gamat1}
N(y,Q^2) = \int d \omega \,e^{-\omega y + \gamma(\omega) \ln Q^2} =
\int d \omega e^{\omega ( - y + C \ln Q^2 )}\,\,, 
\end{equation}
 which means that $N(y,Q^2) $ depends only  on one variable \cite{BL} 
\begin{equation} \label{eq:lev:satsceq}
 - y + C \ln Q^2 = \ln(Q^2_s(x)/Q^2) = \xi \,\,.
\end{equation}
It turns out that $C = \frac{\pi}{4 N_c \alpha_S}$ in double log
approximation \cite{BL,IM}.

\subsection{ Scaling solution:}
Eq.~\ref{eq:lev:satsceq} gives the equation for the critical line $Q^2 =
Q^2_s(x)$ in Fig.~\ref{fig:lev:satpic}.A  more refined  approach to the
solution of the master non-linear equation,  allows us to understand how far
 we have to move from the critical line to see this new scaling. It turns
out that the scaling is not applicable  \cite{LT} only in a narrow (at low $x$ ) band along
the critical line with the width $\ln (Q^2_s(x)/Q^2) \approx 1/4
\alpha_S\,\ln(1/x) \,\ll\,1$ for $  \alpha_S\,\ln(1/x) \,\gg\,1$.

\begin{figure}[h]
\begin{center}
\begin{tabular}{cc}
\epsfig{file=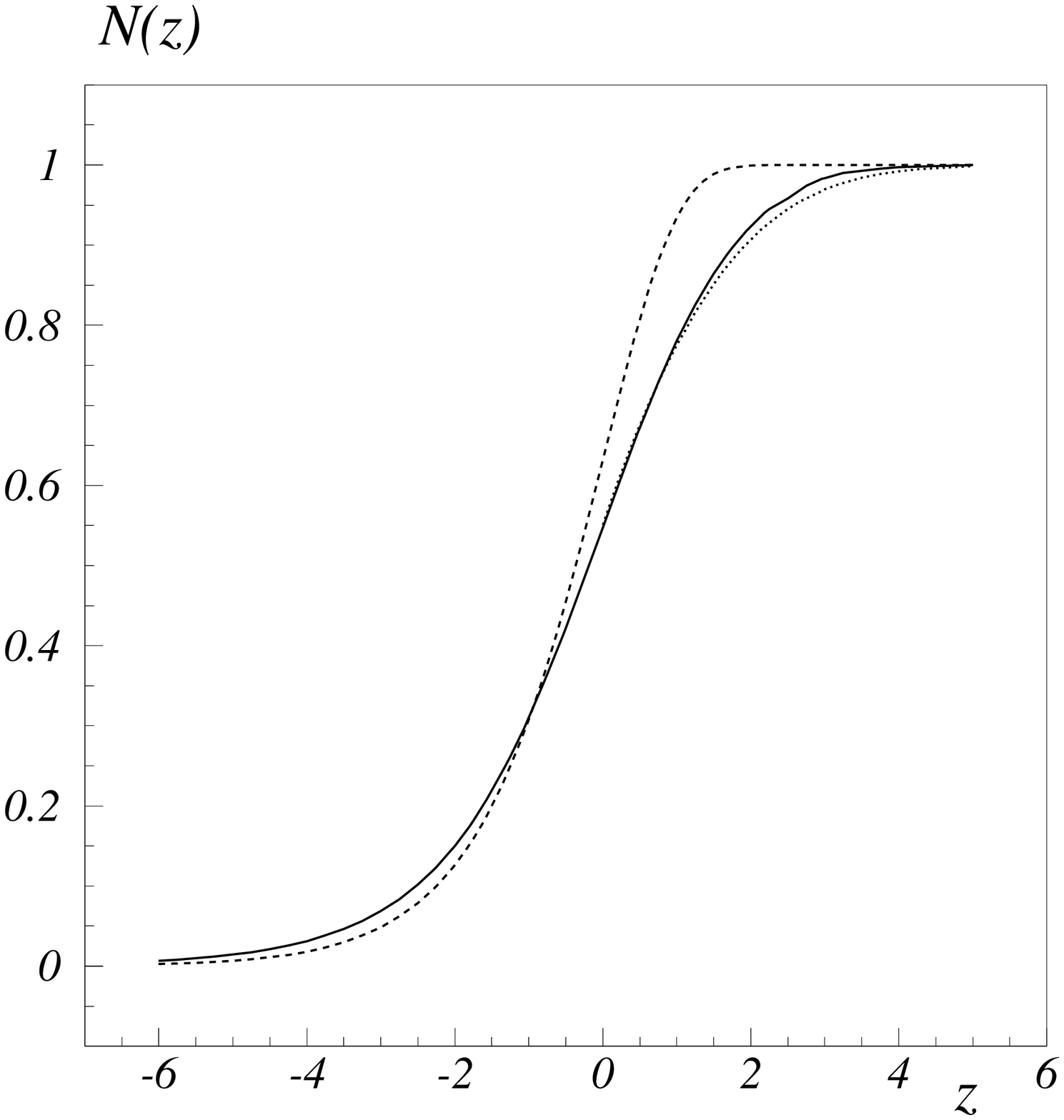,width=8cm,height=8cm}&
\epsfig{file=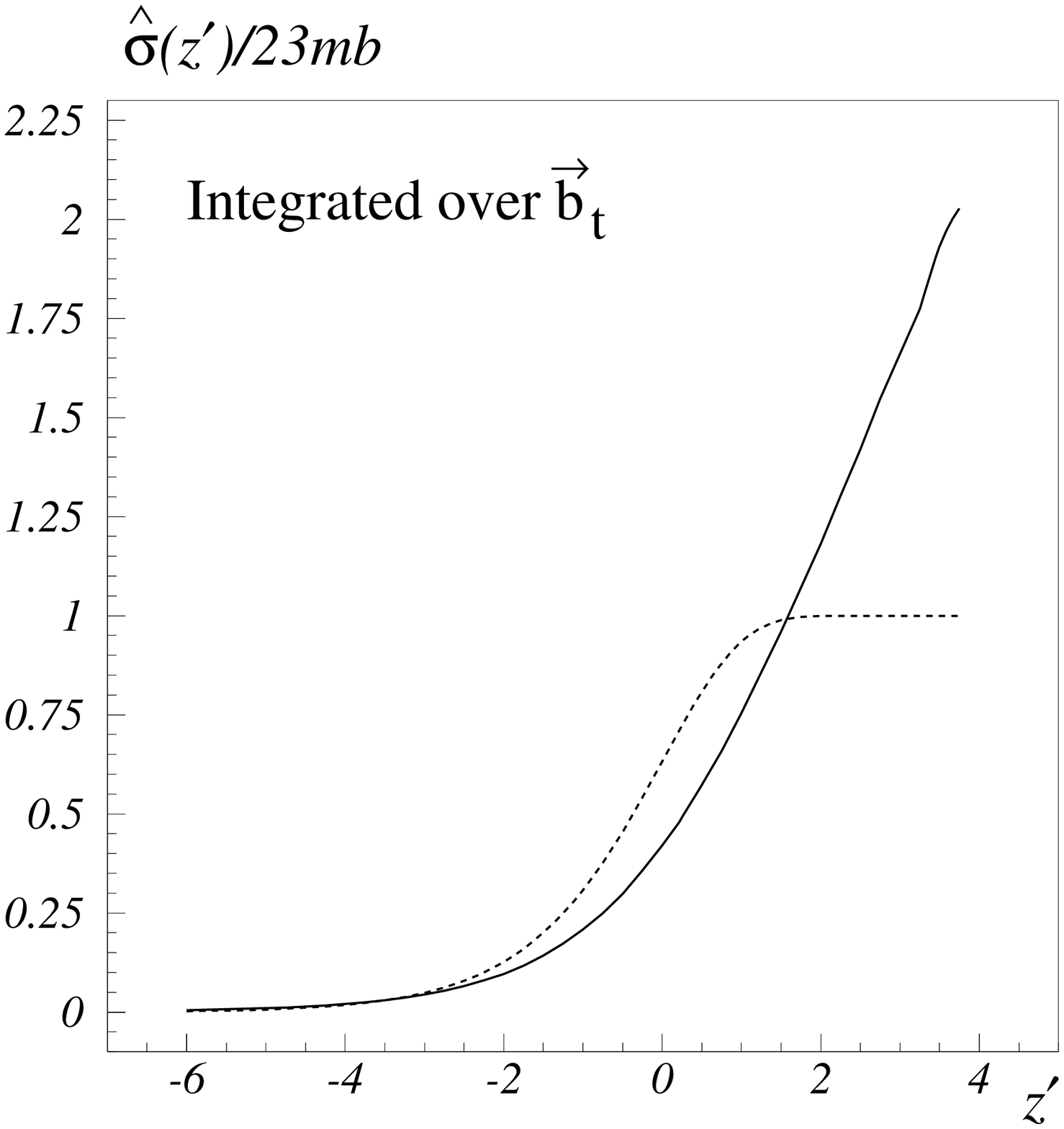,width=8cm,height=8cm}\\
 a)& b) \\                
\end{tabular}
\end{center}
\caption{ (a)  Dipole -- target scattering amplitude $N(z)$ at $b_t=0$  and
(b) dipole -- target cross section $\hat\sigma(z')$ in the
scaling approximation versus scaling variable (a)
$z=\ln(r^2_{\perp}\,Q^2_s(x))$ and (b) $z' = \ln(Q^2_s(x)/Q^2)$: Solid
line is the scaling solution of the master equation, dashed line is
a Golec-Biernat -- Wusthoff model as explained in text and dotted line is
the $z \gg  1$ asymptotic calculated in the first paper of Ref\protect\cite{LT}.}
\label{fig:lev:satsol}
\end{figure}

It should be stressed that two quite different approaches: one is the
analytic solution of the master equation \cite{LT} and the second is the
solution of Wilson renormalization group equation for generating function
\cite{IM}, lead to the same answer and the same picture for the new scaling.
The scaling solution is given in Fig. ~\ref{fig:lev:satsol}. the difference
between $N(z,b_t=0)$ and $\sigma$ is due to the integration over impact parameter $b_t$.

\subsection{ Scaling violation:}
\begin{figure}[h]
\begin{center}
\epsfig{file=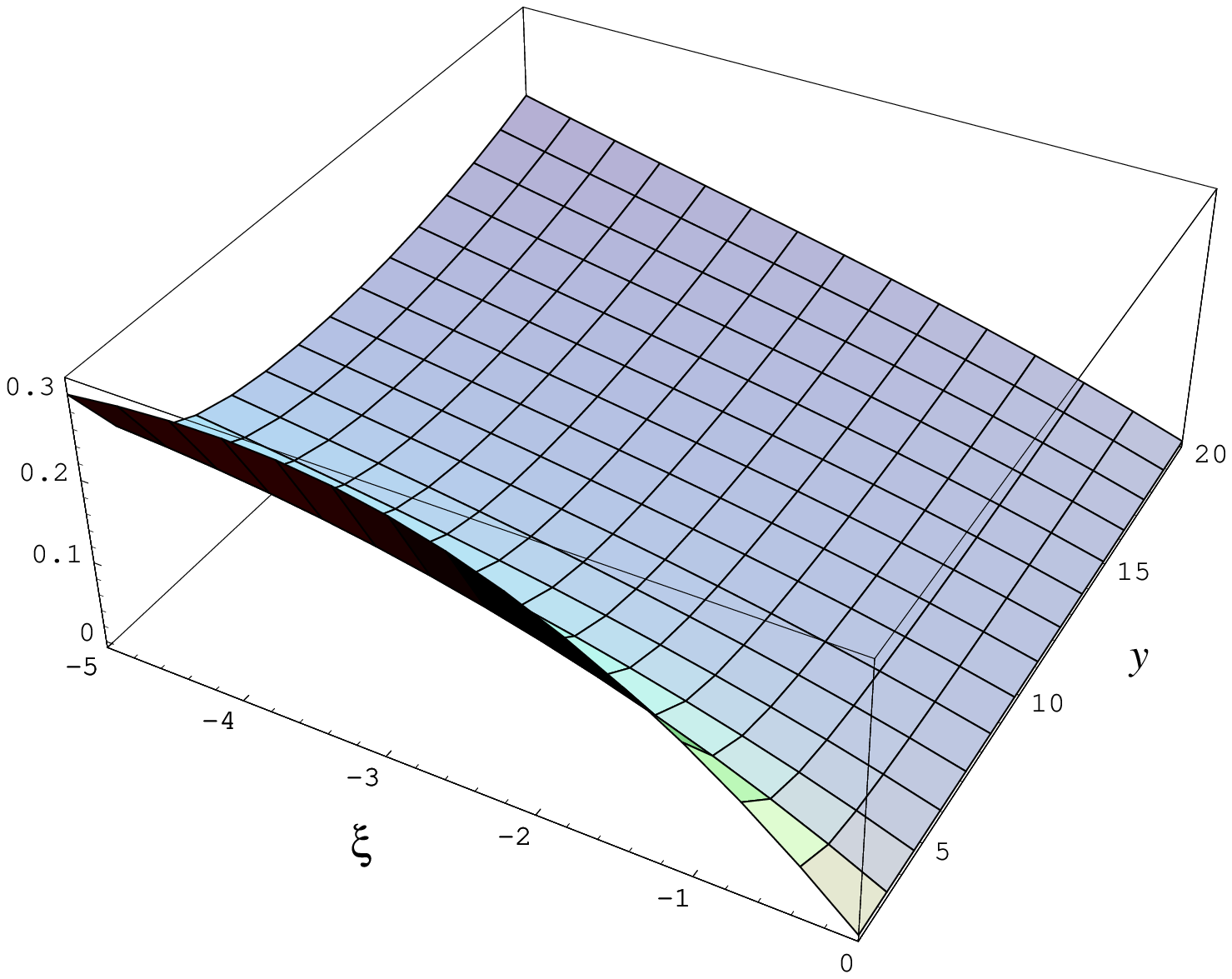, width=10cm,height=10cm}
\end{center}
\caption{ The ratio $\delta N(y,\xi)/ N(\xi)$.}
\label{fig:lev:scvio}
\end{figure}

The question which we wish to address  is how large is the  scaling violation for $Q^2
>Q^2_s(x)$.  In Ref.\cite{LT} one can find first estimates of this scaling
violation (see Fig.~\ref{fig:lev:scvio}

\begin{figure}[h]
\begin{minipage}{10.0cm}
\begin{center}
\epsfxsize=9.4cm
\leavevmode
\hbox{\epsffile{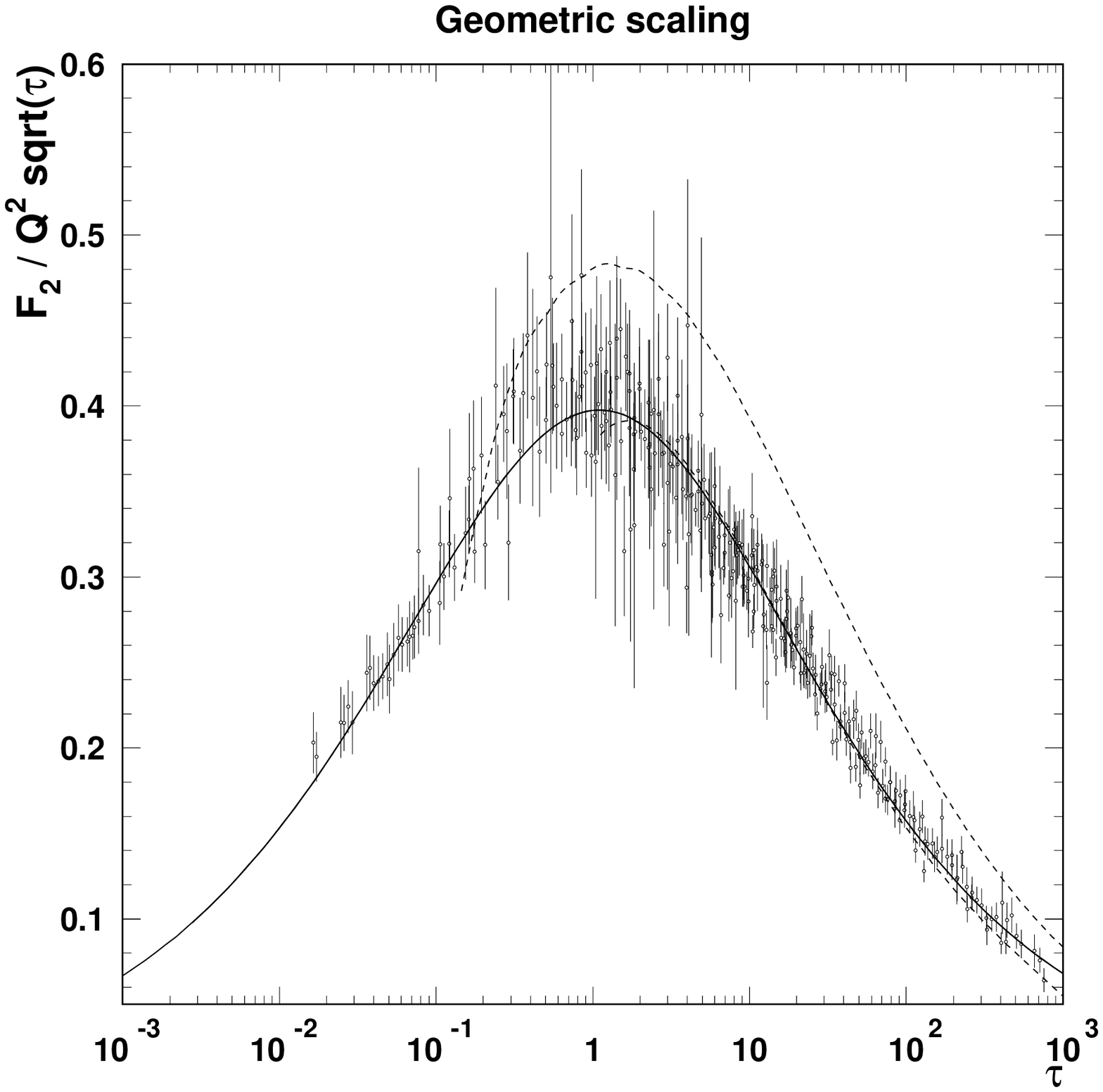}}
\end{center}
\end{minipage}
\begin{minipage}{6.0cm}
\caption{ The HERA data, plotted as a function of $\tau =
Q^2/Q^2_s(x) $. The solid curve is the prediction of the Wusthoff and
Golec-Biernat saturation model, which was calculated  by K. Golec-Biernat
specially for THERA WS.
The two dashed curves give
the prediction of the DGLAP evolution (parameterization GRV'98) for the
THERA kinematic region.}
\label{fig:lev:scG}
 \end{minipage}
\end{figure}

One can see that at all reasonable values of $y = \ln(1/x)$ and 
$\xi = \ln(Q^2_s(x)/Q^2)$ the violation is less than 30$\%$. It means
  that if we 
 want to have  a tool to measure $Q^2_s(x)$ as a value of $Q^2$ at which we see
a scaling violation, we have to establish a scale to an accuracy of  less than 10$\%$.
We think that it is instructive to examine Fig.~\ref{fig:lev:scG} from this
point of view. One can see from this figure that the accuracy of new scaling
phenomena in HERA data is low, and we cannot use these data to measure the value of the
saturation scale.

\section{Parton densities at THERA and LHC energies}

\subsection{Analytic estimates:}
All three theoretical analyses\cite{MU99,LT,IM} of the master equation give
 the same result for the dipole-target amplitude $N(x,r_{\perp};b_t)$, namely
in  the region of low $x$ $ N(x,r_{\perp};b_t) \,\longrightarrow\, 1$ in
 accordance with the unitarity constraint. Eq.~\ref{eq:lev:O2} leads to gluon
structure function which can be calculated  as follows
\begin{eqnarray}  \label{eq:lev:xgsat}
xG(x,Q^2) \,&=&\,\frac{8}{\pi^3}\int^1_x \,\frac{d
    x'}{x}\,\int^{\infty}_{4/Q^2}\,\frac{d
    r^2_{\perp}}{r^4_{\perp}}\,\int\,d^2\,b_t \,N(x', r_{\perp};b_t)
  \nonumber \\
 & \longrightarrow & \frac{2}{\pi^2}\,Q^2\,R^2\,\int^{y_{cr}}_{y} dy' \,\,,
\end{eqnarray}
where $y_{cr}$ is the solution of the equation $Q^2_s(y_{cr})
\,=\,Q^2$ and $R$ is the size of the target.
 Therefore, the answer for the asymptotic is clear,  and  it  can be
rewritten in the form \cite{MU99,LT,IM}:
\begin{equation}  \label{eq:lev:xgsatli}
xG(x,Q^2)\,\,\longrightarrow\,\,\frac{N^2_c - 1}{4 \pi
  N_c}\,\frac{1}{\alpha_S}\,\,R^2\,Q^2\,
\ln(Q^2_s(x)/Q^2)\,\,.
\end{equation}
In Eq.~\ref{eq:lev:xgsatli} we assumed that the $b_t$ -distribution does not
depend on $x$. Actually this  is not correct  and we have some shrinkage of the
diffraction peak \cite{LT}. All numerical coefficients in
Eq.~\ref{eq:lev:xgsatli} are fixed in the double log approximation and have
to be checked to a  better accuracy.

\subsection{Numerical results:}
\begin{figure}[h]
\begin{tabular}{c c}
$\mathbf{x \,= \,10^{-7} }$ & $\mathbf{x\,=\,10^{-6} }$ \\
 \epsfig{file=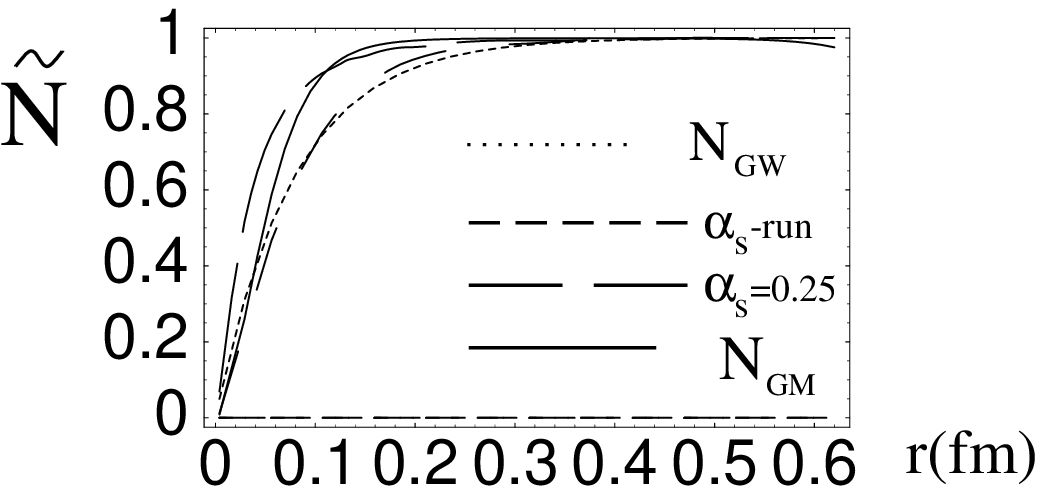,width=82mm, height=40mm}&
\epsfig{file=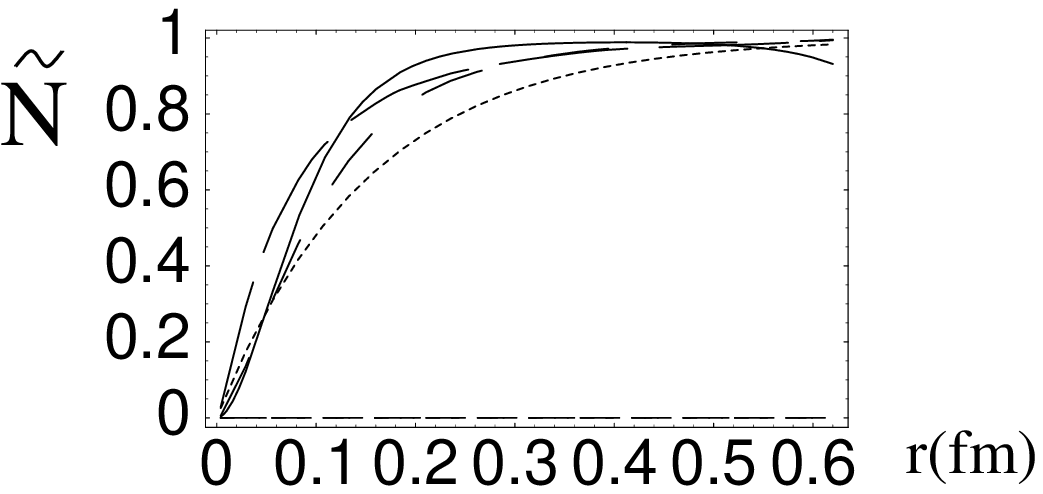,width=82mm, height=40mm}\\
$\mathbf{x\,=\,10^{-4} }$ & $\mathbf{x\,=\,10^{-3} }$ \\
 \epsfig{file=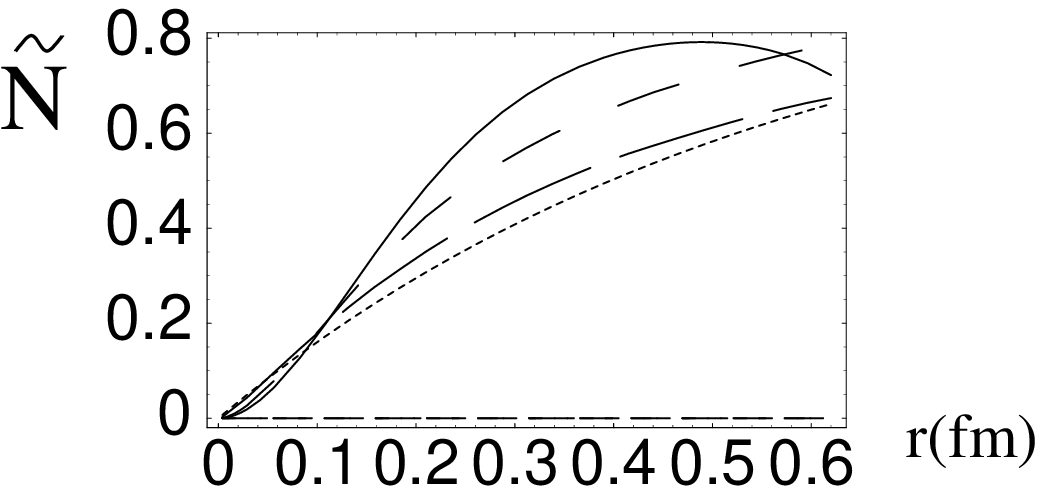,width=82mm, height=40mm}&
\epsfig{file=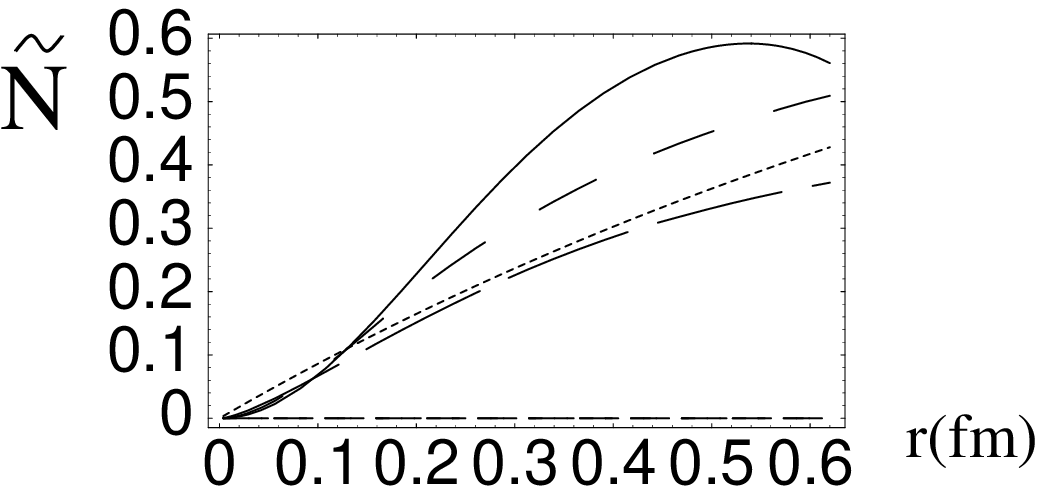,width=82mm, height=40mm}\\
\end{tabular}
\caption{  The comparison between the solutions $\tilde N$,
Glauber-Mueller formula  \protect\cite{GLMAM},
 and Golec-Biernat and Wuesthoff (GW) model. The four curves correspond to two
different
solutions $\tilde N_{\alpha_S=0.25}$ (large dashes),
 $\tilde N_{\alpha_S - running}$(small dashes),
$N_{GM}$ (continuous line), and $N_{GW}$ of the GW model (dots).}
\label{fig:lev:nsol1}
\end{figure}

The analytic approach leads to an  understanding, but also gives a certain check
of a numerical procedure for  solving the master equation. At the moment, we
have two attempts to solve the equation numerically \cite{BR,GLMNL} which can
illustrate the  effect that  the non-linear evolution will have  for the
extrapolation   of HERA data to higher energies ( lower $x$).
 (see Fig.~\ref{fig:lev:nsol1}.  

\begin{figure}[h]
\begin{tabular}{c c}
\epsfig{file=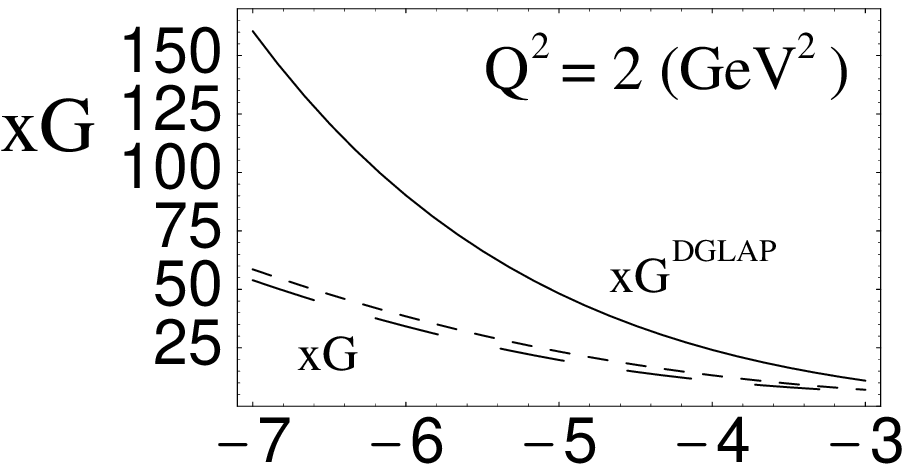,width=80mm, height=50mm}&
\epsfig{file=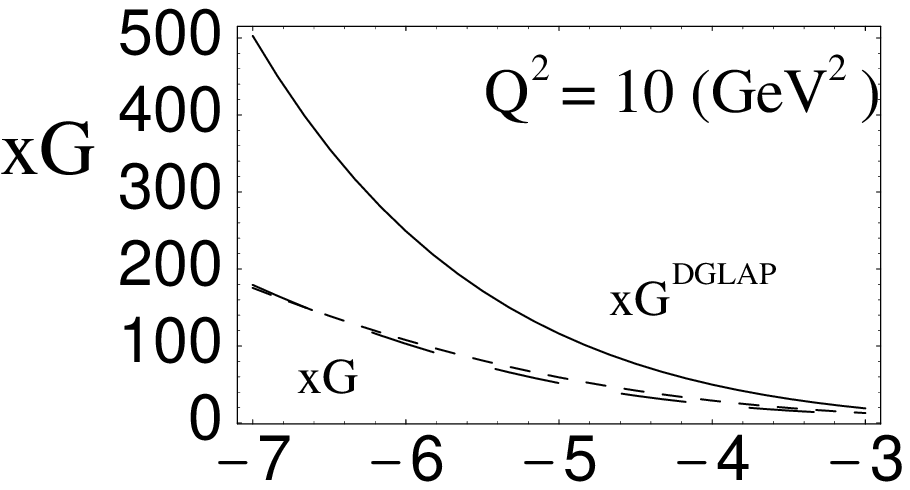,width=80mm, height=50mm}\\
 \epsfig{file=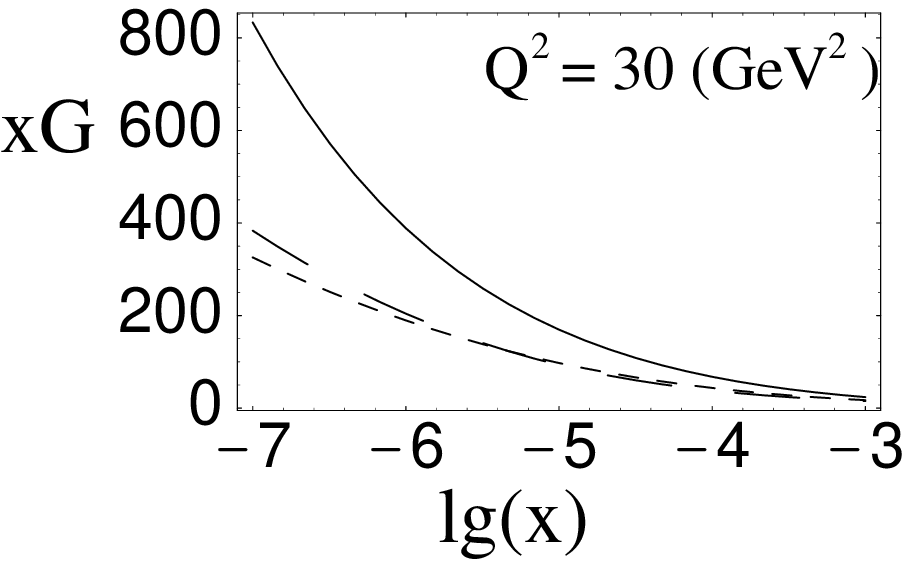,width=80mm, height=50mm}&
\epsfig{file=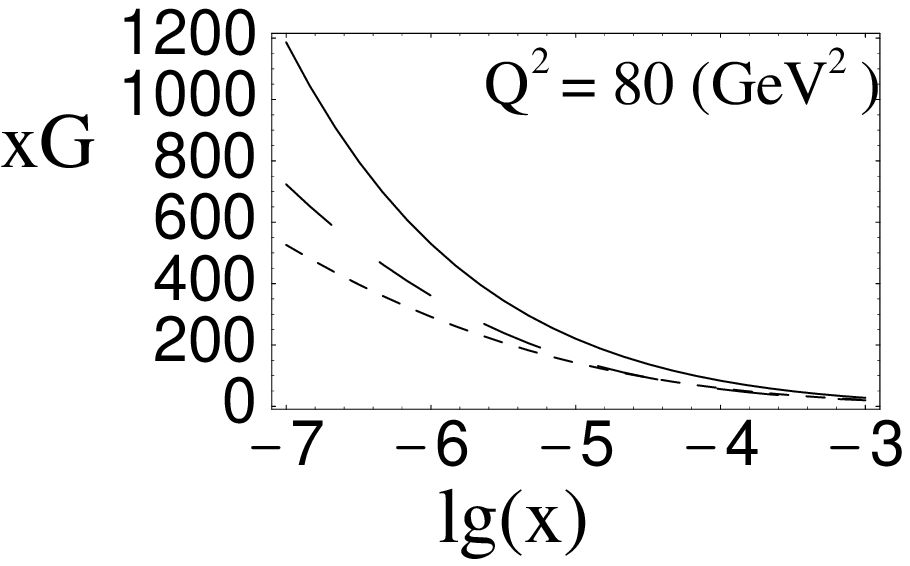,width=80mm, height=50mm}\\
\end{tabular}   
 \caption{ The function $xG$ is plotted versus $\lg (x)$.
 The small dashes
  correspond to the solution with the running $\alpha_S$, while the large 
dashes
are used for  the constant $\alpha_S=0.25$. The continuous line is the GRV
parameterization.}
\label{fig:lev:nsol2}
\end{figure}

We can learn at least two lessons: first, the approximate models cannot be
 used for the predictions at higher energies; and  second, the saturation
  phenomenon
could be rather strong both at THERA and LHC energies. Such a strong
 modification  of the $r_{\perp}$ - behaviour due to non-linear corrections
reflects in $Q^2$ and $x$ behaviour of gluon density as one  can see in Fig. ~\ref{fig:lev:nsol2}

Fig.~\ref{fig:lev:nsol2} shows that the taming of the parton densities growth
included in the master equation can easily reduce the value of the gluon
density at THERA and LHC energies by  2$\div$ 3 times, at different values of
$Q^2$.

\section{ Summary}
\subsection{From the first principles:}
The main message that I want to deliver in this presentation is very simple:
{\bf \it Due to the hard work of number of experts, the theory of high parton  density QCD has been established},
and this theory includes
the proof of :
\begin{itemize}
\item \quad The nonlinear evolution equation for the dipole-target amplitude
  at fixed $b_t$ ($Im\, a(x,r_{\perp};b_t) = N(x,r_{\perp};b_t)$);
\item \quad The saturation of the parton densities as  $x \rightarrow 0$ which
  means that $N(x,r_{\perp};b_t)\,\longrightarrow\,1$ at low $x$;
\item \quad The new scaling phenomena $N(x,r_{\perp};b_t)\,=\,N( r^2_{\perp} \cdot
  Q^2_s(x),b_t)$ for $Q^2\,<Q^2_s(x)$;
\item \quad The sharp increase of the saturation scale $Q^2_s(x)$ in the
  region of low $x$;
\item \quad The importance of the saturation effects in taming the  growth 
  of the gluon density at THERA and LHC energies.
\end{itemize}
We would like to stress that during  past two  decades we have developed
the perturbative QCD methods\cite{KOV},  based on the correct degrees of
  freedom: colour dipoles at high energies\cite{DOF3}
and  created  new operator methods of tackling  this problem,
  such as the Wilson Loop Operator Expansion\cite{BA} and the effective Lagrangian
  approach\cite{MV,ILM,IM}. My personal feeling is that the future is in the
  operator methods since they provide  a possibility to describe 
  the matching of the high density QCD with real non-perturbative QCD with
  large coupling constant    in a unique way. However, I would like to emphasize the positive
  aspects of the pQCD  approach : a clear physical picture  in 
  the pQCD calculations and, because of this picture, the transparent  
  understanding of the meanings of observables which we calculate in pQCD. 

The whole development of this new area of theory I consider as a triumph of
approach of my teacher, Prof. Gribov, to high energy physics, which could be
formulated as {\it \bf  ``Physical picture - first, mathematics -  after if  ever''}\cite{DOK}.
 \,  Indeed, I am very certain that the remarkable progress, that we see now, was
 possible only, because the picture that has been discussed in the first
 section, was correct from the beginning.

 \subsection{ Our   prejudices: }
I decided to finish my paper listing the  prejudices (some of them ) that I have fought 
during the two decades:
\begin{itemize}
\item \quad The DGLAP evolution equation is more fundamental and has better
  proof than non-linear one. It is not true because 
\begin{itemize}
\item \quad  The GLAP equation yields 
  the parton densities which  violate the unitarity constraints;
\item \quad The next to leading order corrections to the DGLAP evolution
  equation have to be included at energies $\ln (1/x) \,\geq 1/\alpha^2_S $ (
  see section 3.4.1) which are much higher that the energy $\ln (1/x) \,\geq
  \ln(1/\alpha_S)/\alpha_S$
when the parton rescatterings and recombinations start to be essential;
\item \quad
 It  does
  not contain any proof of why the  higher twist contribution is smaller, than the
  leading twist one,  which
  has been incorporated in these equations; 
\item \quad These equations demand
  initial conditions which we do not know how to treat theoretically or
  extract experimentally.
\end{itemize} 
\item \quad By choosing the initial value for the DGLAP evolution $Q^2_0$ to be
  large enough,  we can claim that the higher twists contributions are small.
It was shown that the anomalous dimension of the higher twists is much larger
that the leading one \cite{BHT,LRHT} and at any value of $Q^2$ at low $x$, the higher twist
will exceed  the leading twist (DGLAP) contribution. Therefore, the practical
procedure of solving   the DGLAP evolution equation, cannot be considered to
be 
free of justified  criticism ;
\item \quad The calculation of higher order corrections of pQCD will improve
  our description of the experimental data. The pQCD series are the
  asymptotic series and they have intrinsic  accuracy,  which  cannot be
  improved by calculating  additional  orders in $\alpha_S$,
and this accuracy is rather poor in the region of low $x$. To evaluate
errors we have to use the next order calculation in pQCD as an error, but with
the same initial conditions.
\end{itemize}

\subsection{The map of QCD}

\begin{figure}[h]
\begin{minipage}{10.0cm}
\begin{center}
\epsfxsize=9.4cm
\leavevmode
\hbox{\epsffile{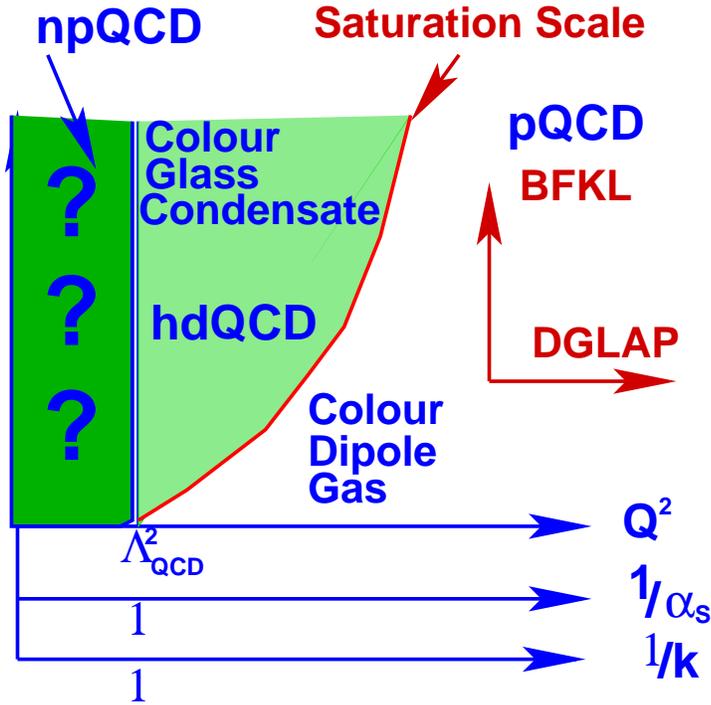}}
\end{center}
\end{minipage}
\begin{minipage}{6.0cm}
\caption{ The map of QCD. $\alpha_S$ is the running QCD coupling constant,
  $\kappa =  (3\pi^2 \alpha_S/2 Q^2) \cdot (xG(x,Q^2)/\pi R^2)$ is the
packing factor for the gluons. The saturation scale $Q^2 = Q^2_s(x)$ is the
line on which $\kappa = 1$.}
\label{fig:lev:map}
 \end{minipage}
\end{figure}      

Fig.~\ref{fig:lev:map} displays the QCD map  which is as the result of this
approach.  Three separate regions are distinguished by the size of the
variables $Q^2$ and $x$. Each region corresponds to quit different physics. 

{\bf 1. Perturbative QCD domain} where the constituents are of small size and
their density is rather small so that we can neglect their interactions
(recombinations). The packing factor $\kappa \,< 1 $ , and we can view
this region as the traditional region for the linear QCD evolution equations
( the DGLAP and BFKL ones ). In this region the parton system appears as a Colour
Dipole Gas, since we have here a rather dilute system of colour dipoles.

{\bf 2. High parton density  QCD domain} in which the constituents are still
small and we can use weak coupling methods, but their density is so large that
the gluon field in this region becomes strong. We  call the parton system
in this domain a colour glass condensate\cite{LMLL}, because gluons have
colour, their fields evolve very slowly relative to the natural scale and
are disordered. In simple words glass is a solid on a short time scale and a
liquid on long time scale. In lectures by L. MacLerran \cite{LMLL}the reader can find
the arguments why the parton system  in this domain looks like a glass.
We have a condensate because of high parton density and strong gluon fields
in  this region.

{\bf 3. Non perturbative QCD domain} in which the QCD coupling is large
($\alpha_s \,\,\approx\,\,1$). In this region the confinement of quarks and
gluons occurs and a real non perturbative approach should be developed
here. The key words for this domains are QCD vacuum, lattice QCD, effective
Lagrangian approach, QCD - string theory duality and other lofty words about
possible solutions of non perturbative QCD.

\subsection{ Acknowledgments :}

I would like to thank all participants of low $x$ WG at THERA WS . Hot and useful
discussions with them on high density QCD problems provided  a stimulus  for writing
this presentation. I am very grateful  for illuminating discussions to 
comrades  in arms:  Ian Balitsky,Jochen Bartels,  Mikhail Braun, Asher Gotsman,
Krystoff Golec-Biernat, Edmond Iancu, Dima Kharzeev, Boris Kopeliovich, Yura Kovchegov,  Alex
Kovner, Ian Kwiecinski, Uri Maor,   Larry
McLerran  with his Minnesota-BNL team , Al Mueller and Heribert Weigert. I
equally thank my students Mikhael Lublinsky,Eran Naftali and Kirill Tuchin
for being good friends and labourers  in the  hard work on low $x$ problems.

I would also like  to express a deep acknowledgment  to my opponents for keeping
me in a  fighting mood. The present status of high density QCD is a
challenge to them and , I hope, they will scrutinize and criticize   our
approach on the same professional level as it has been proven. I am sure that
such a  criticism  will give a new impetus for further development.

I wish to thank all HERA experimentalists for their beautiful data and the
deep interest in  low $x$ problems. I have expressed my point of view on HERA
data in different publication and, here, I wanted to give a review of the
theoretical scene for them. I am very much  indebted  to them  because HERA data
revived  a question on Pomeron structure , my  first and strongest love which
will never pass.

 This research was supported in part by the BSF grant $\#$
9800276, by the GIF grant $\#$ I-620-22.14/1999
  and by
Israeli Science Foundation, founded by the Israeli Academy of Science
and Humanities.


\begin{thebibliography}{99} 
\bibitem{DGLAP}

V.N. Gribov and L.N. Lipatov,{\it Sov. J. Nucl. Phys.} {\bf 15} (1972) 438
;\\
 L.N. Lipatov, {\it Yad. Fiz.} {\bf 20}(1974)  181 ;\\
G.  Altarelli
and
G. Parisi,   {\it Nucl. Phys.} {\bf B126} (1977) 298;\\
 Yu.L.
Dokshitser,       {\it
Sov. Phys. JETP} {\bf 46}(1977) 641.      
\bibitem{BFKL}
E.A. Kuraev, L.N. Lipatov and V.S. Fadin: {\it Sov. Phys. JETP} {\bf 45}
(1978) 199;\\
Ya. Ya. Balitsky and L.N. Lipatov: {\it Sov. J. Nucl. Phys.} {\bf 28}
(1978) 22.            
\bibitem{GLR}
L.V. Gribov, E.M.  Levin 
 and M.G.  Ryskin, {\it Phys. Rep}
{\bf 100} (1983) 1; {\it Nucl.Phys.} {\bf B188} ((1981) 555. 
                                                                  
\bibitem{MUQI}
A.H. Mueller and J. Qiu, {\it Nucl. Phys.} {\bf B268} (1986) 427.
\bibitem{MV}
L. McLerran and R. Venugopalan,{\it Phys. Rev. } {\bf D49} (1994)
2233,3352, {\bf 50} (1994) 2225, {\bf 53} (1996) 458, {\bf 59} (1999)
094002. 
\bibitem{MU99}
A.H. Mueller, {\it Nucl. Phys.} {\bf B558} (1999) 285.
\bibitem{BJ}
R.P. Feyman, {\it `` Photon - Hadron Interaction''}, Benjamin, NY,1972;\\
J.D. Bjorken, J. Kogut and D.E. Soper, {\it Phys. Rev.} {\bf D3} (1971)
1382;\\
J.D. Bjorken, {\it Phys. Rev.} {\bf D1} (1970) 1376;\\
J.D. Bjorken and E. A. Paschos,  {\it Phys. Rev.} {\bf 185} (1969) 1975.
\bibitem{BK}
J. Bartels and H. Kowalski, {\it `` Diffraction at HERA and the Confinement
  Problem''}, DESY-00-154,{\tt hep-ph/0010345}.
\bibitem{DOF1}
A. Zamolodchikov, B. Kopeliovich and L. Lapidus, {\it JETP Lett.} {\bf 33}
(1981) 595;
\bibitem{DOF2}
E.M.   Levin and M.G.  Ryskin, {\it Sov. J. Nucl. Phys.} {\bf 45} (1987)
150.
\bibitem{DOF3}
A. H. Mueller, {\it Nucl. Phys.} {\bf B335} (1990) 115.
 \bibitem{MU94}
A.H.  Mueller, {\it  Nucl. Phys.}  {\bf B415} (1994) 373.
\bibitem{AGL}
A.L. Ayala, M.B. Gay Ducati and E.M. Levin, {\it Nucl. Phys.}{\bf B493}(1997)
305, {\bf B511} (1998) 355;
\bibitem{NZ}
N.N. Nikolaev and B.G. Zakharov, {\it Z. Phys.} {\bf C49} (1991) 607;\\
E.M. Levin, A.D. Martin, M.G.  Ryskin and T. Teubner, {\it Z. Phys.} {\bf
C74} (1997) 671.
\bibitem{KM}
Yu. V. Kovchegov and L. McLerran, {\it Phys. Rev.} {\bf D60}
(1999) 054025;\\
Yu. V. Kovchegov and E. Lervin, {\it  Nucl.Phys.}  {\bf B577}
(2000) 221.                                                                                 
\bibitem{BHT}
J. Bartels, {\it Z.Phys.} {\bf C60} (1993) 471, {\it Phys. Lett.} {\bf B298}
(1993) 204.
\bibitem{LRHT}
E.M. Levin, M.G. Ryskin and A.G. Shuvaev, {\it Nucl. Phys.} {\bf B387} (1992)
589.
\bibitem{LLS}
E. Laenen,E. Levin and A.G. Shuvaev, {\it Nucl. Phys.} {\bf B419} (1994) 39.
\bibitem{LL}
E. Laenen and E. Levin, {\it Nucl. Phys.} (1995) 207; {\it
  Ann.Rev.Nucl.Part.Sci.} {\bf 44} (1994) 199-246. 
\bibitem{BA}
I. Balitsky, {\it Nucl. Phys.} {\bf B463} (1996) 99, {\it ``High energy QCD
  and Wilson lines''}, {\tt hep-ph/0101042 },JLAB-THY-00-44.
\bibitem{KOV}
Yu. V. Kovchegov, {\it Phys. Rev.} {\bf D60} (2000) 034008.
\bibitem{BR}
M. Braun, {\it Eur. Phys.J.} {\bf C16} (2000) 337.
\bibitem{3P}
A. Mueller and B. Patel, {\it Nucl. Phys.} {\bf B425} (1994) 471;\\
J. Bartels and M. Wuesthoff, {\it Z. Phys.} {\bf C66} (1995) 157.
\bibitem{ILM}
E. Iancu, A. Leonidov and L. McLerran, {\it ``Nonlinear Gluon Evolution in
  the Color Glass Condensate''}, BNL-NT-00/24,{\tt hep-ph/0011241};
{\it ``The renormalization group equation for the color glass condensate"},
BNL-NT-01-3, {\tt hep-ph/0102009}.
\bibitem{IM}
E. Iancu and L. McLerran, {\it `` Saturation and Universality in QCD at small
  $x$''}, SACLAY-T01-026, {\tt hep-ph/0103032}.
\bibitem{EFLA}
J. Jalilian-Marian, A. Kovner, L. McLerran  and  H.
Weigert, {\it Phys. Rev.}{\bf D55}(1997) 5414;\\
J. JuliLian-Marian, A. Kovner and  H.
Weigert, {\it Phys. Rev.} {\bf 59}(1999) 014015;\\
J. Jalilian-Marian, A. Kovner, A.
Leonidov and  H. Weigert, {\it Phys. Rev.} {\bf 59}(1999) 014014,034007,\\
A. Kovner, J.Guilherme Milhano and  H. Weigert,
{\it Phys. Rev.} {\bf D62}(2000)      114005,\\
 H. Weigert, {\it ``Unitarity at small Bjorken $x$''} NORDITA-2000-34-HE,
{\tt hep-ph/0004044}.                                                           
\bibitem{MUKO}
Yu. V. Kovchegov and A.H. Mueller, {\it Nucl. Phys.} {\bf B529} (1998) 451.
\bibitem{GLMAM}
E. Gotsman et al. {\it ``Has HERA reached a new QCD regime?''}, DESY-00-149,
{\tt hep-ph/0010198}.
\bibitem{AGLFRT}
A.L. Ayala, M.B. Gay Ducati and E.M. Levin, {\it Phys. Lett.} {\bf B388}
(1996) 188.
\bibitem{LT}
E. Levin and K. Tuchin, {\it Nucl. Phys.} {\bf B573} (2000) 833; {\it `` New
  scaling aty high energy DIS''}, TAUP-2659-2000,{\tt  hep-ph/0012167 }; {\it
  Nonlinear evolution and saturation for heavy nuclei in DIS''},
TAUP-2664-2001,{\tt hep-ph/0101275}.
\bibitem{KOV1}
Yu.V. Kovchegov, {\it Phys. Rev.} {\bf D61} (2000) 074018.
\bibitem{COLBAR}
J.C. Collins and J. Kwiecinski, {\it Nucl. Phys.} {\bf B335} (1990) 89;\\
J. Bartels, J. Blumlien and G. Shuler, {\it Z. Phys.} {\bf C50} (1991) 91.
\bibitem{GW}
K. Golec-Biernat and M. Wuesthoff, {\it `` Diffractive parton distributions
  from the saturation model''},DESY-00-180,{\tt hep-ph/0102093};
{ \it Phys.Rev.} {\bf D60} (1999) 114023;{\bf D59} (1999) 014017. 

\bibitem{GLMNL}
M. Lublinsky et al., {\it `` Non-linear evolution and parton distributions at
  LHC and THERA energies''}, TAUP-2667-2000,{\tt hep-ph/0102321}.
\bibitem{BL}
J. Bartels and E. Levin, {\it Nucl.Phys.} {\bf B387} (1992) 617. 
\bibitem{SGK}
A. Stasto, K. Golec-Biernat and J. Kwiecinski,{ \it Phys.Rev.Lett.} {\bf 86}
(2001) 596.
\bibitem{LMLL}
L. McLerran, {\it`` The Color Glass Condensate and Small $x$ Physics: 4
  Lectures''}, {\tt hep-ph/0104285}.
\bibitem{DOK}
Yu. Dokshitzer, {\it ``V.N. Gribov: 1930 -1997''},  {\tt physics/9801025}.
\end{thebibliography}
\end{document}